\documentstyle[11pt,aaspp4]{article}
\begin{document}

\title{DIRECT Distances to Nearby Galaxies Using Detached Eclipsing
Binaries and Cepheids.  VIII. Additional Variables in the Field M33B
Discovered with Image Subtraction\footnote{Based on observations
obtained with the 2.1m telescope at the Kitt Peak National
Observatory.}}

\author{B. J. Mochejska, J. Kaluzny}
\affil{Copernicus Astronomical Center, Bartycka 18, 00-716 Warszawa}
\affil{\tt e-mail: mochejsk@camk.edu.pl, jka@camk.edu.pl}
\author{K. Z. Stanek, D. D. Sasselov\altaffilmark{2} 
\& A. H. Szentgyorgyi}
\affil{Harvard-Smithsonian Center for Astrophysics, 60 Garden St.,
Cambridge, MA~02138}
\affil{\tt e-mail: kstanek@cfa.harvard.edu, sasselov@cfa.harvard.edu
aszentgyorgyi@cfa.harvard.edu}
\altaffiltext{2}{Alfred P. Sloan Research Fellow.}

\begin{abstract}

DIRECT is a project to obtain directly the distances to two Local
Group galaxies, M31 and M33, which occupy a crucial position near the
bottom of the cosmological distance ladder.

As the first step of the DIRECT project we have searched for detached
eclipsing binaries (DEBs) and new Cepheids in the M31 and M33 galaxies
with 1m-class telescopes. In this eighth paper we present a catalog of
variable stars discovered in the data from the followup observations
of DEB system D33J013337.0+303032.8 in field M33B $[(\alpha,\delta)=
(23.\!\!\arcdeg48, 30.\!\!\arcdeg57), {\rm J2000.0}]$, collected with
the Kitt Peak National Observatory 2.1m telescope. In our search
covering an area of $108\arcmin^2$ we have found 895 variable stars:
96 eclipsing binaries, 349 Cepheids, and 450 other periodic, possible
long period or non-periodic variables. Of these variables 612 are
newly discovered. Their light curves were extracted using
the ISIS image subtraction package. For 77\% of the variables we
present light curves in standard $V$ and $B$ magnitudes, with the
remaining 23\% expressed in units of differential flux.

We have discovered a population of first overtone Cepheid candidates
and for six of them we present strong arguments in favor of this
interpretation. 

The catalog of variables, as well as their photometry (about
$9.2\times 10^4$ BV measurements) and finding charts, is available
electronically via {\tt anonymous ftp} and the {\tt World Wide Web}.
The complete set of the CCD frames is available upon request.
\end{abstract}
\keywords{binaries: eclipsing -- Cepheids -- distance scale --
galaxies: individual (M33) -- stars: variables: other}

\section{Introduction}

Starting in 1996 we undertook a long term project, DIRECT
(i.e. ``direct distances''), to obtain the distances to two important
galaxies in the cosmological distance ladder, M31 and M33. These
``direct'' distances will be obtained by determining the distance of
Cepheids using the Baade-Wesselink method and by measuring the
absolute distance to detached eclipsing binaries (DEBs). While the
cosmological distance scale has been the subject of numerous recent
observation campaigns, especially those enabled by the Hubble Space
Telescope (HST) and massive variability studies of the Magellanic
clouds, M33 has not been re-surveyed since the photographic
survey of Kinman, Mould \& Wood (1987).

M31 and M33 are the stepping stones to most of our current effort to
understand the evolving universe at large scales.  First, they are
essential to the calibration of the extragalactic distance
scale. Second, they constrain population synthesis models for early
galaxy formation and evolution and provide the stellar luminosity
calibration. There is one simple requirement for all this---accurate
distances. These distances are now known to no better than 10-15\%, as
there are discrepancies of $0.2-0.3\;{\rm mag}$ between various
distance indicators (e.g.~Huterer, Sasselov \& Schechter 1995; Holland
1998; Stanek \& Garnavich 1998).

DEBs have the potential to establish distances to M31 and M33 with an
unprecedented accuracy of better than 5\% and possibly to better than
1\%. Detached eclipsing binaries (for reviews see Andersen 1991;
Paczy\'nski 1997) offer a single step distance determination to nearby
galaxies and may therefore provide an accurate zero point calibration
of various distance indicators -- a major step towards very accurate
determination of the Hubble constant, presently an important but
daunting problem for astrophysicists. DEBs have been recently used to
obtain accurate distance estimate to the Large Magellanic Cloud
(e.g. Guinan et al.~1998; Udalski et al.~1998).

The detached eclipsing binaries have yet to be used as distance
indicators to M31 and M33. According to Hilditch (1996) there was only
{\em one} eclipsing binary of any kind known in M33 (Hubble 1926). The
recent availability of large-format CCD detectors and inexpensive CPUs
has made it possible to organize a massive search for periodic
variables, which will produce a handful of good DEB candidates. These
can then be spectroscopically followed-up with the powerful new 6.5-10
meter telescopes.

The study of Cepheids in M33 has a venerable history (Hubble 1926).
Freedman, Wilson \& Madore (1991) obtained multi-band CCD photometry
of some of the Cepheids discovered in photographic surveys, to build a
period-luminosity relations in M33. However, the sparse photometry and
the small sample (11 Cepheids) do not provide a good basis for
obtaining direct Baade-Wesselink distances (see, e.g., Krockenberger,
Sasselov \& Noyes 1997) to Cepheids---the need for new digital
photometry has been long overdue.

As the first step of the DIRECT project we have searched for DEBs and
new Cepheids in the M31 and M33 galaxies. We have analyzed five
$11\arcmin\times11\arcmin$ fields in M31, A-D and F (Kaluzny et
al. 1998, 1999; Mochejska et al. 1999; Stanek et al. 1998, 1999;
hereafter Papers I, IV, V, II, III). A total of 410 variables, mostly
new, were found: 48 eclipsing binaries, 206 Cepheids and 156 other
periodic, possible long-period or non-periodic variables. We have
analyzed two fields in M33, A and B (Macri et al. 2001a; hereafter
Paper VI) and found 544 variables: 47 eclipsing binaries, 251 Cepheids
and 246 other variables.

As a second step, we started followup observations of selected DEBs in
both the M31 and M33 galaxies with larger telescopes in order to
construct more precise and well sampled light curves for them. As a
by-product of the monitoring of DEB system D33J013346.2+304439.9 with
the Kitt Peak National Observatory 2.1m telescope, we have extracted
434 variable stars in field M33A: 63 eclipsing binaries, 305 Cepheids,
and 66 other variables, of which 280 were newly discovered (Mochejska
et al. 2001; hereafter Paper VII). We have also discovered eight bona
fide first-overtone Cepheid candidates. 

In this paper, eighth in the series, we present a catalog of variable
stars found in the same field as the detached eclipsing binary
D33J013337.0+303032.8 using followup observations collected at the
Kitt Peak National Observatory 2.1m telescope. The paper is organized
as follows: Section 2 provides a description of the observations. The
data reduction procedure is outlined in Section 3. The catalog of
variable stars is presented in Section 4, followed by its brief
discussion in Section 5. Section 6 deals with the first overtone
Cepheid candidates. The concluding remarks are stated in Section 7.

\section{Observations}

\begin{figure}[t]
\plotfiddle{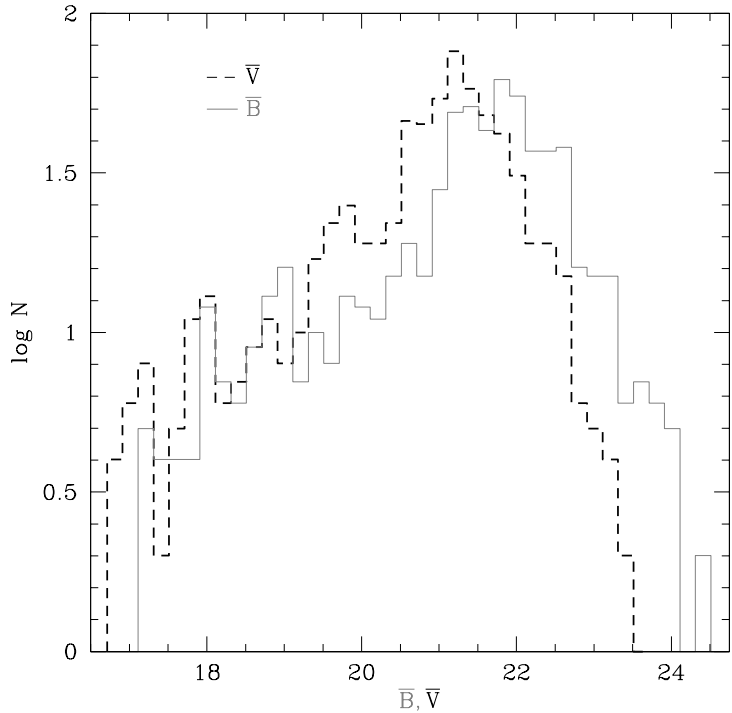}{8.8cm}{0}{50}{50}{-150}{-85}
\caption{Distributions in B (continuous line) and V (dashed line)
of variable stars in the field M33B.}
\label{fig:dist}
\end{figure}

The data discussed in this paper was obtained at the Kitt Peak
National Observatory\footnote{Kitt Peak National Observatory is a
division of NOAO, which are operated by the Association of
Universities for Research in Astronomy, Inc. under cooperative
agreement with the National Science Foundation.} 2.1m telescope
equipped with a Tektronix $2048\times2048$ CCD (T2KA camera) having a
pixel scale $0.305\arcsec/pixel$ during two separate runs, from
September 29th to October 5th, 1999 and from November 1st to 7th,
1999. The primary observing targets were three detached eclipsing
binaries, one in each of the fields M33A, M33B and M31A, discovered
previously as part of the DIRECT project (Papers VI and II). For field
M33B we collected $74\times600s$ exposures in the $V$ filter and
$30\times600s$ in the $B$ filter.\footnote{The complete list of
exposures for this field and related data files are available through
{\tt anonymous ftp} on {\tt cfa-ftp.harvard.edu}, in {\tt
pub/kstanek/DIRECT} directory. Please retrieve the {\tt README} file
for instructions.  Additional information on the DIRECT project is
available through the {\tt WWW} at {\tt
http://cfa-www.harvard.edu/\~\/kstanek/DIRECT/}.} The exposure times
varied slightly to compensate for the changes of seeing
conditions. The typical seeing was $1.\!\!\arcsec5$. The field was
observed through airmass ranging from 1 to 1.9, with the average at
1.2. The completeness of our data starts to drop rapidly at about 21.2
mag in $V$ and 21.8 mag in $B$, judging from the magnitude
distributions of the variable stars (Fig. \ref{fig:dist}).

\section{Data Reduction, Calibration and Astrometry}

During the second night of the first observing run we found that the
camera had a non-linear response. The data was corrected for this
effect using the method described in Paper VII. 

The photometry for the variable stars was extracted using the ISIS
image subtraction package (Alard \& Lupton 1998, Alard 2000a) from the
$V$ and $B$-band data. We followed the same reduction procedure as
described in detail in Paper VII. One difference is that instead of
the ISIS CZERNY subroutine for preliminary period determination we
used a program based on the multiharmonic analysis of variance
algorithm (Schwarzenberg-Czerny 1996).

Due to residual non-linearity, our photometry could not be calibrated
from observations of standard stars. The coefficients for the color
terms of the transformation were derived from the comparison of our
NGC 6791 photometry with the data from the KPNO 0.9m telescope
(Kaluzny \& Udalski 1992). The offsets were determined relative to 735
stars above $V=20$ mag from the DIRECT catalog of stellar objects in
M33 (Macri et al. 2001b). The following transformations were adopted:
\begin{eqnarray*}
  v = V - 5.501 + 0.039\cdot(B-V)\\
b-v =  0.146 + 0.927\cdot(B-V)\\
  b = B - 5.355 - 0.034\cdot(B-V)
\end{eqnarray*}
The instrumental $V$ and $B$-band light curves of the variables were
transformed to the standard system by adding the appropriate offsets,
as the coefficients next to the color terms are very small.

Equatorial coordinates were determined for the $V$ template star list,
expanded with the variables with no $V$-band photometry. The
transformation from rectangular to equatorial coordinates was derived
using 894 transformation stars with $V<19.5$ from the DIRECT catalog
of stellar objects in M33 (Macri et al. 2001b). The average difference
between the catalog and the computed coordinates for the
transformation stars was $0.\arcsec06$ in right ascension and
$0.\arcsec06$ in declination. 

\section{Catalog of Variables}
\subsection{Classification}

The variables we are most interested in are Cepheids and eclipsing
binaries (EBs). We therefore searched our sample of variable stars
primarily for these two classes of variables. The variable stars were
preliminarily classified as eclipsing, Cepheid or miscellaneous by
visual inspection, based on the shape of their light curves. The
variables for which neither $V$ nor $B$-band magnitude could be
determined (with only flux light curves or having periods in excess of
14 days) were not reclassified further. 

In order to obtain as clean a sample of Cepheids as possible, we have
inspected the location of the Cepheid variable candidates on a $V/B-V$
CMD. All of the Cepheid candidates having highly discrepant colors
were reclassified as other periodic variables. The candidates with
light curves in only one of the bands were checked on the
period-luminosity relation for that band. Extreme outliers were also
moved to the other periodic variable category.

The EBs with light curves expressed in magnitudes for at least one
band, were fitted with a model described in Papers I and II. Within
our assumption the light curve of an EB is determined by nine
parameters: the period, the zero point of the phase, the eccentricity,
the longitude of periastron, the radii of the two stars relative to
the binary separation, the inclination angle, the fraction of light
coming from the bigger star and the uneclipsed magnitude. Eclipsing
binary candidates for which a satisfactory fit was not achieved, were
reclassified as other periodic variables.

\begin{figure}[p]
\plotfiddle{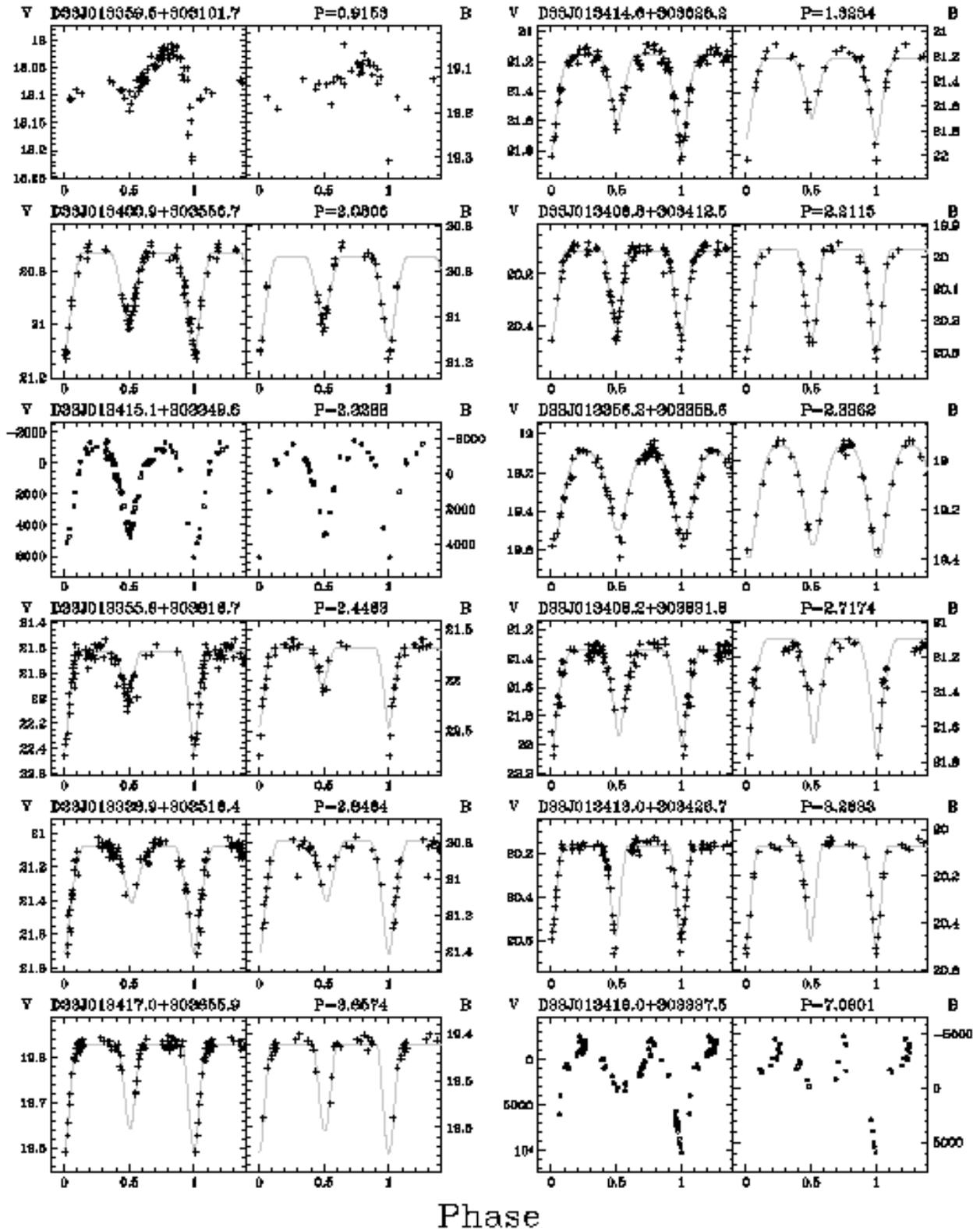}{19.5cm}{0}{88}{88}{-280}{-75}
\caption{Selected $BV$ light curves of eclipsing binaries found in the
field M33B. The points on the light curves expressed in magnitudes are
marked with crosses, the flux ones by open circles. The thin
continuous line represents the best fit model for each star and
photometric band.}
\label{fig:ecl}
\end{figure}

In the following sections \ref{sub:ecl}-\ref{sub:misc} we present the
parameters and light curves for the 895 identified variable
stars.\footnote{The $BV$ photometry and $V$ finding charts for all
variables are available from the authors via the {\tt anonymous ftp}
from the Harvard-Smithsonian Center for Astrophysics and can be also
accessed through the {\tt World Wide Web}.}  All tables are sorted by
increasing period, unless otherwise noted. Variables for which more
than one plausible period was found are indicated by colons appended
to their listed periods. We adopt a naming convention after Macri et
al. (2001a) based on the J2000.0 equatorial coordinates, in the
format: D33J$hhmmss.s$+$ddmmss.s$. The first three fields ($hhmmss.s$)
correspond to right ascension expressed in hours, the last three
($ddmmss.s$) to declination, expressed in degrees, separated by the
declination sign. As none of the newly discovered variables are
present in previous variable star catalogs, we refer the reader to
Tables 5 and 6 in Paper VI for cross-identifications.

\subsection{Eclipsing Binaries}
\label{sub:ecl}

We have found a total of 96 eclipsing binaries in field M33B.  In
Table \ref{tab:ecl} we present the parameters for the 89 EBs with a
magnitude light curve in at least one band. For each variable we list
its name, period P, magnitudes $V_{max}$ and $B_{max}$ of the system
outside of the eclipse, and the radii of the binary components
$R_1,\;R_2$ in the units of the orbital separation.  We also give the
inclination angle of the binary orbit to the line of sight $i$ and the
eccentricity of the orbit $e$.  The reader should bear in mind that
the values of $V_{max},\;B_{max},\; R_1,\;R_2,\;i$ and $e$ are derived
with a straightforward model of the eclipsing system, so they should
be treated only as reasonable estimates of the ``true'' value. Table
\ref{tab:ecl_flux} lists seven EBs with flux light curves only. For
each variable we give its name and period P. Figure \ref{fig:ecl}
presents the phased light curves of 12 sample EBs (see also
Table~\ref{tab:lc_ecl}).

\subsection{Cepheids}
\label{sub:cep}

A total of 349 Cepheid variables were found in field M33B. In Table
\ref{tab:ceph} we present 292 Cepheids with a magnitude light curve in
at least one band. For each variable, we list its name, period P,
flux-weighted average magnitudes $\langle V\rangle$ and $\langle
B\rangle$ and the $V$ and $B$-band amplitudes $A_V$ and $A_B$. Due to
the short time base of our observations, reliable flux-weighted
magnitudes could only be determined for variables with periods shorter
than 14 days. 

\begin{figure}[p]
\plotfiddle{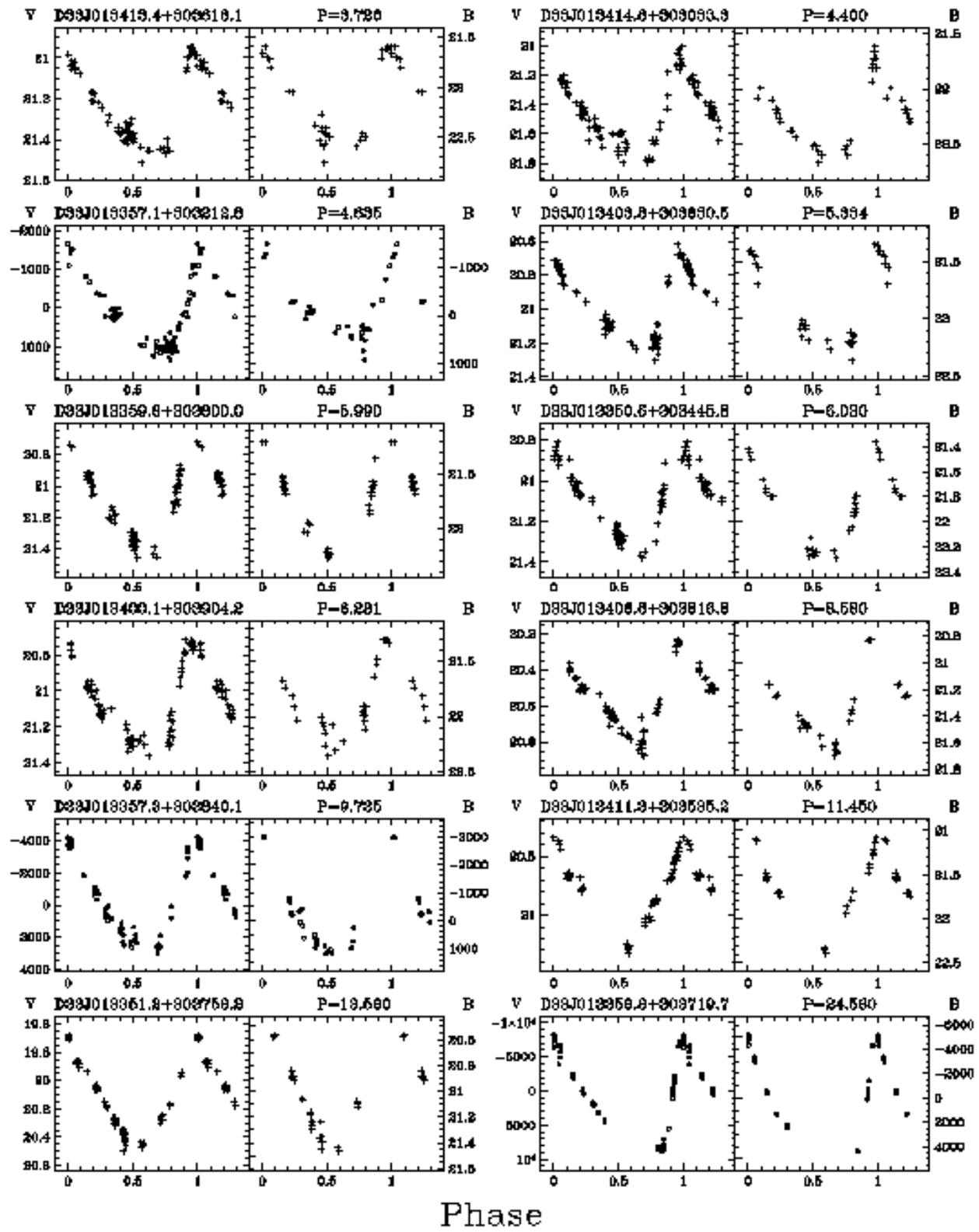}{19.5cm}{0}{88}{88}{-280}{-75}
\caption{Same as Fig. \ref{fig:ecl}, but for Cepheid variables.}
\label{fig:ceph}
\end{figure}

We have extracted light curves for 37 Cepheids with
longer periods, all of them identified previously in Paper VI and made
them available via {\tt anonymous ftp}. In Table \ref{tab:ceph_flux}
we list 20 Cepheids with flux light curves only. For each variable we
list its name and period P. Figure \ref{fig:ceph} presents the phased
light curves of 12 sample Cepheids (see also Table
\ref{tab:lc_ceph}).

\begin{figure}[ht]
\plotfiddle{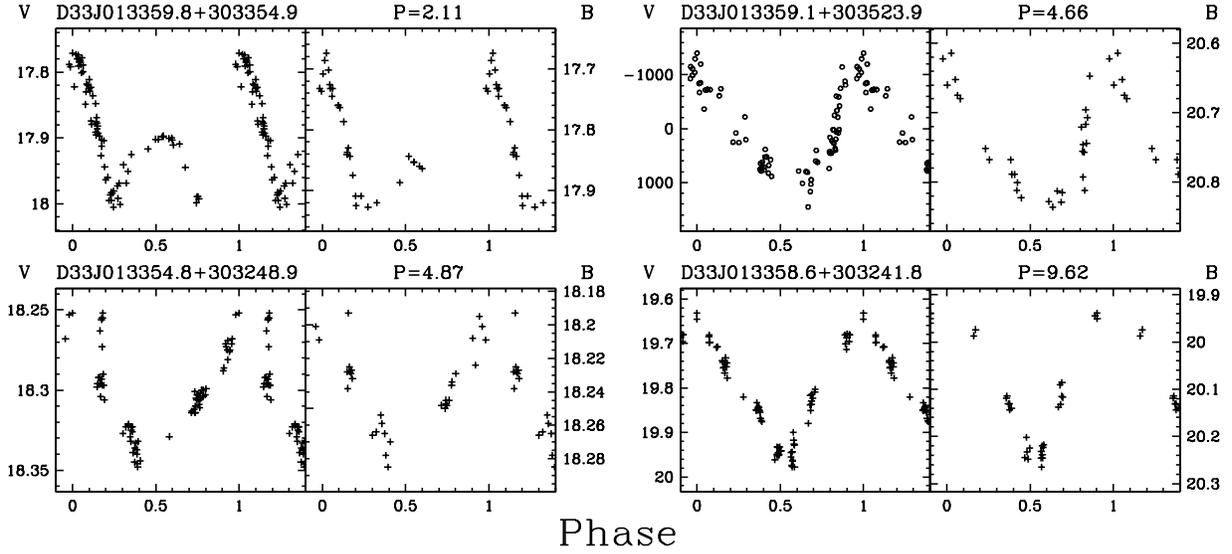}{6.5cm}{0}{88}{88}{-280}{-450}
\caption{Same as Fig. \ref{fig:ecl}, but for other periodic variables.}
\label{fig:oth}
\end{figure}

\begin{figure}[ht]
\plotfiddle{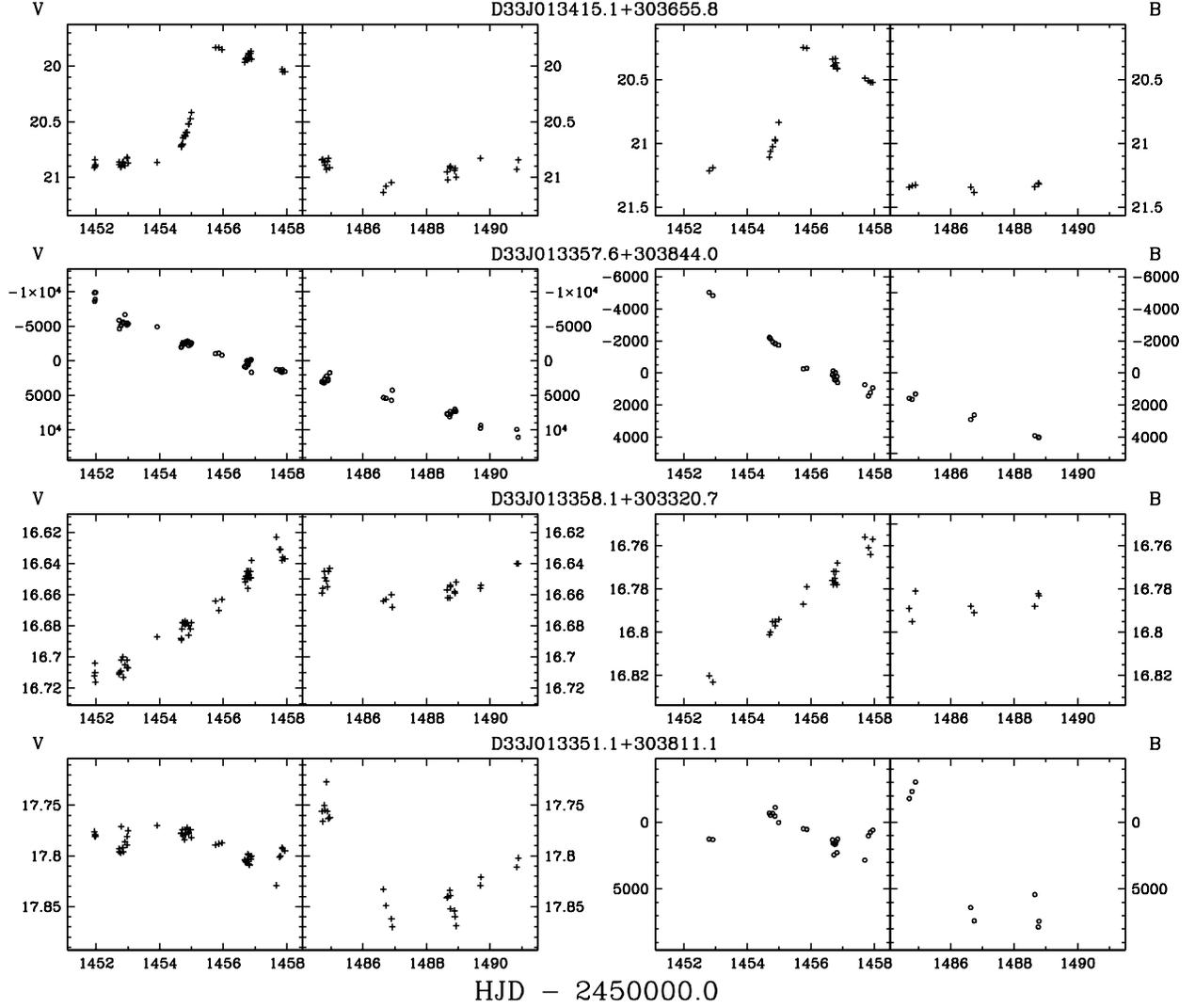}{13cm}{0}{88}{88}{-280}{-75}
\caption{Same as Fig. \ref{fig:ecl}, but for miscellaneous variables.}
\label{fig:misc}
\end{figure}

\subsection{Other Periodic Variables}
\label{sub:per}
In Table \ref{tab:per} we present the parameters of 30 periodic
variables.  For each variable we list its name, period P, the
magnitudes $V$ and $B$ and the $V$ and $B$-band amplitudes $A_V$ and
$A_B$. The $V$ and $B$ columns list the magnitudes outside of the
eclipses $V_{max}$ and $B_{max}$ for the eclipsing variables and
flux-weighted average magnitudes $\langle V\rangle$ and $\langle
B\rangle$ for the other variables. We have also found 29 other
variables with periods of the order of 50-100 days (identified
previously in Paper VI), which we do not list in the table. Their
light curves are available via {\tt anonymous ftp}. In
Table~\ref{tab:lc_per} we list the light curves of all periodic
variables. Figure \ref{fig:oth} presents the phased light curves of 4
sample other periodic variables.

\subsection{Miscellaneous Variables}
\label{sub:misc}
In Table \ref{tab:misc} we present the parameters of 391 miscellaneous
variables.  For each variable we list its name, the average magnitudes
$\bar{V}$ and $\bar{B}$ and the $V$ and $B$-band amplitudes $A_V$ and
$A_B$. About 25\% of those variables are most likely periodic, with
periods in excess of 14 days. In Table~\ref{tab:lc_misc} we list the
light curves of all miscellaneous variables. Figure \ref{fig:misc}
presents the light curves of four sample miscellaneous variables.

\begin{figure}[ht]
\plotfiddle{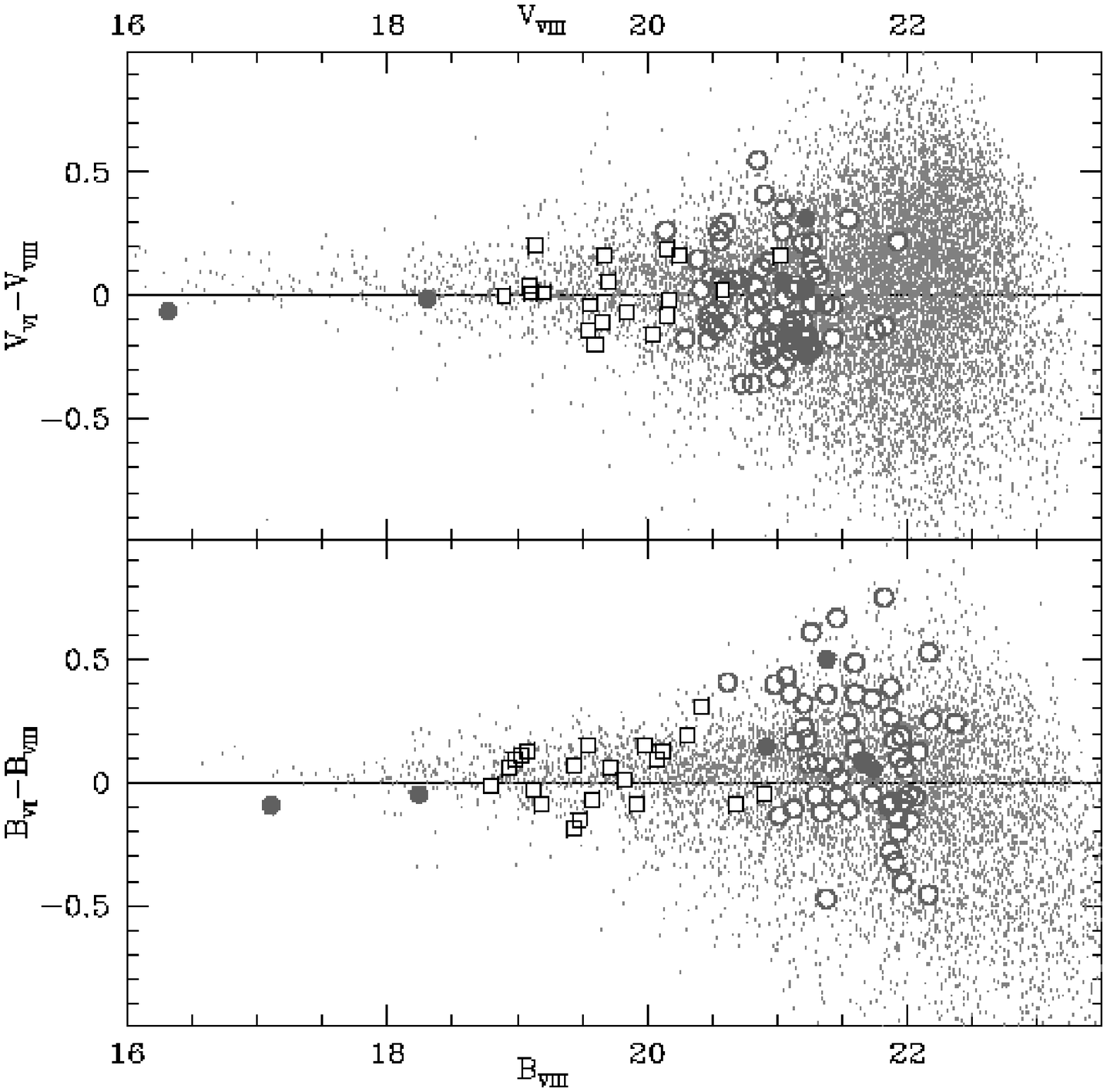}{9.7cm}{0}{55}{55}{-180}{-100}
\caption{A comparison between variable star photometry in the $V$
(upper panel) and $B$-band (lower panel) between our catalog and Paper
VI. EBs are denoted by squares and Cepheids by circles. All other
stars in the field are plotted with small dots in the background for
reference.}
\label{fig:cmp}
\end{figure}

\subsection{Comparison with the catalog in Paper VI}
\label{sect:cmp}
\begin{figure}[ht]
\plotfiddle{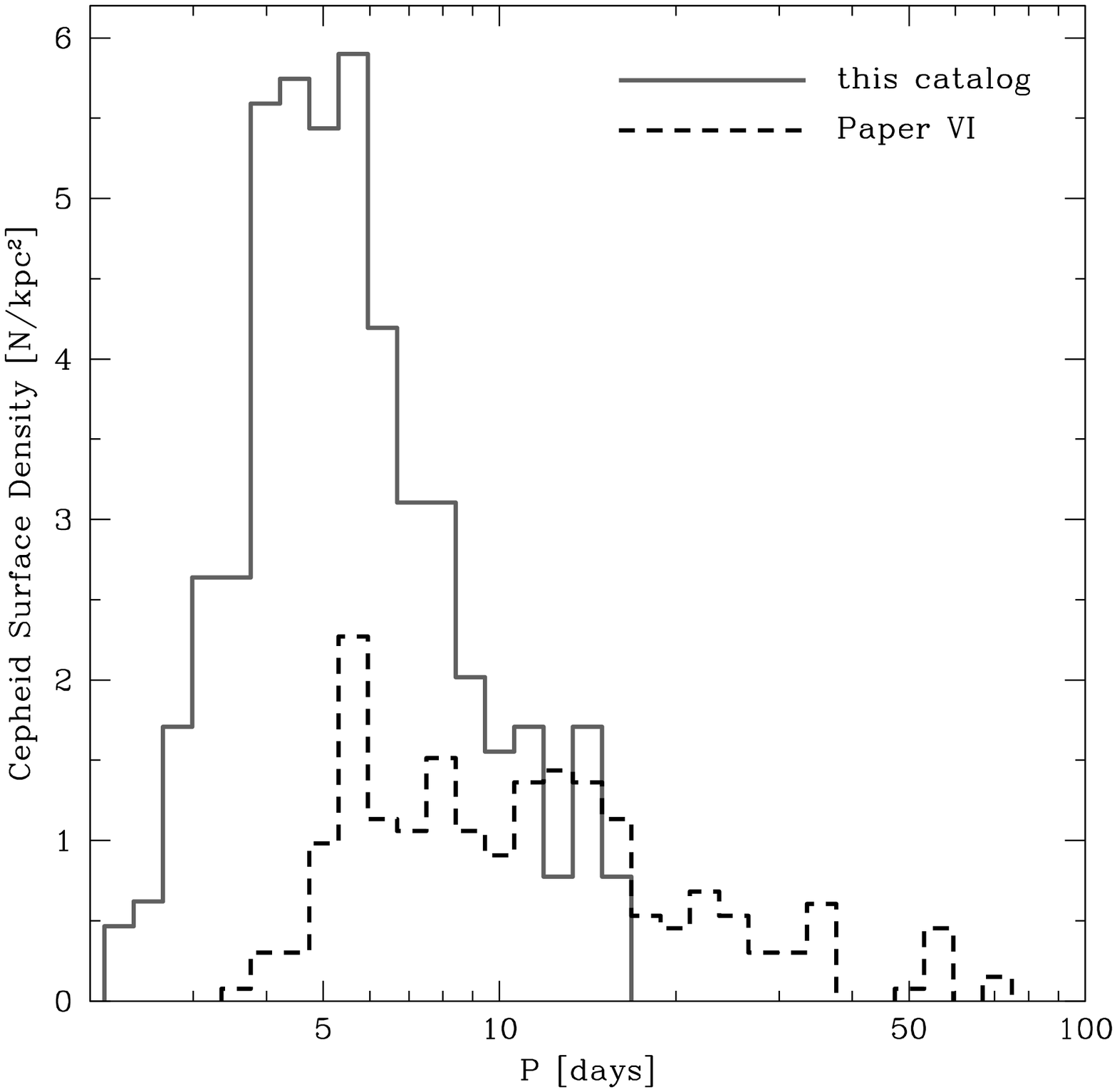}{9.7cm}{0}{55}{55}{-180}{-95}
\caption{A histogram of the surface density of Cepheids (N/arcmin$^2$)
as a function of their period. The solid line represents all Cepheids
from our catalog with $P<14$ days, the dashed line -- the Cepheids
from Paper VI.}
\label{fig:per}
\end{figure}

In Figure \ref{fig:cmp} we compare our $V$ and/or $B$ photometry for
23 EBs (squares), 76 Cepheids (open circles) and periodic variables
(filled circles) with the values listed in the DIRECT catalog of
variables in M33 (Paper VI). As reference we plot in the background
similar comparisons for all the stars in the field (dots). The
variables, with some exceptions, have $\Delta B$ and $\Delta V$
distributions roughly similar to the rest of the stars, although they
do show a slight tendency to be fainter in our catalog.  This trend is
especially prominent in the $V$-band for the EBs, but is absent in
their $B$-band comparison. There are also a few faint Cepheids, which
show unusually large differences in photometry, of the order of
0.6-0.8 mag. In paper VII we have inspected the light curves of
several such variables in both catalogs (see Fig. 8 therein). It seems
that these discrepancies are caused in large part by the fact that
fixed position photometry is prone to identify and fit a profile at
the supplied position even when the star is below the detection
threshold, resulting in the false measurement of a fainter magnitude.

\begin{figure}[ht]
\plotfiddle{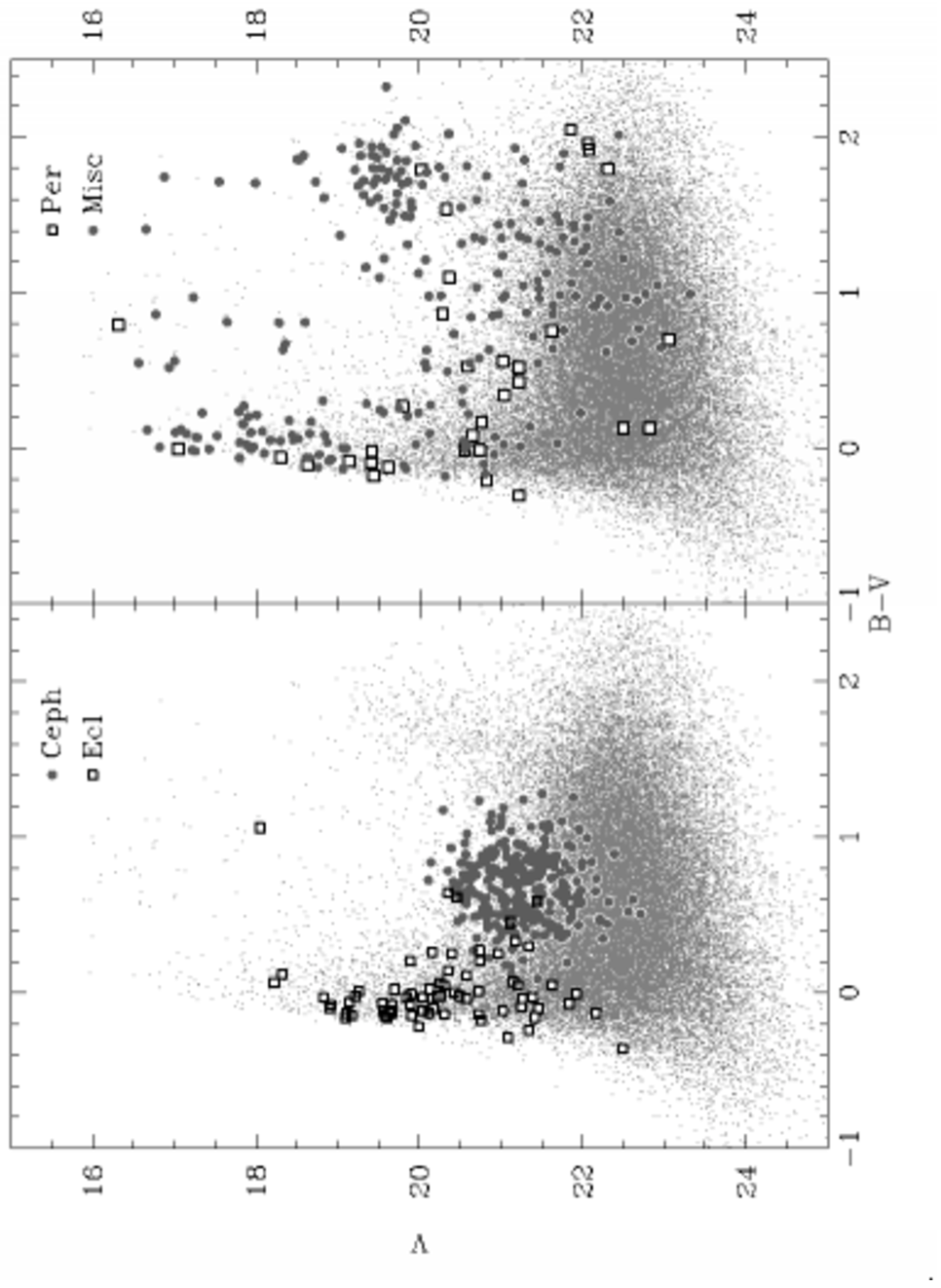}{8.5cm}{-90}{100}{100}{-420}{430}
\caption{The $V/B-V$ CMD for the variable stars in field M33B. The
EBs (open squares) and Cepheids (filled circles) are shown in 
the left panel, periodic (filled squares) and miscellaneous 
variables (open circles) in the right panel.}
\label{fig:cmd}
\end{figure}

In Figure \ref{fig:per} we plot a histogram of the surface density of
Cepheids (N/kpc$^2$) as a function of their period, assuming a
distance of 840 kpc to M33 (Freedman, Wilson \& Madore 1991). The
solid line represents all Cepheids from our catalog with $P<14$ days,
the dashed line -- the Cepheids from Paper VI. The areas covered in
this search and in Paper VI are 108 and 222 arcmin$^2$. As our
observations were carried out during two one-week runs spaced one
month apart, we lack the baseline to detect long period
Cepheids. Since our data were collected with an instrument 2.6 times
larger in area and with better seeing, we have a higher detection rate
for short period Cepheids. The Paper VI catalog, due to the much
longer baseline of observations, contains more long period Cepheids.

\section{Discussion}

In Fig. \ref{fig:cmd} we plot the positions of the variable stars on
the $V/B-V$ CMD. The EBs are denoted by open squares and Cepheids 
by filled circles in the left panel, the periodic variables by 
open squares and miscellaneous by filled circles in the right
panel. 

All but a few of the EBs occupy the upper main sequence. Several fall
in the region occupied by Cepheids. An inspection of their light
curves confirmed that these variables are most likely genuine EBs, and
not Cepheids phased with twice their period. These variables may be
suffering from higher than average reddening or from blending with red
stars. One of the EBs, D33J013359.5+303101.7, with $V=18.05$ and
$B-V=1.06$, is located clearly apart from the rest. The light curve of
this variable is shown in Fig. \ref{fig:ecl} without the model
overlaid, for clarity. The lack of a fairly constant maximum light
between the eclipses indicates that it is a contact binary. The
brightness of the system between the secondary eclipse and the primary
shows a slower rise and a steeper decline. The light curve between the
primary eclipse and the secondary is more sparsely sampled, so we have
checked it in our earlier data (Paper VI), and it seems to be more
symmetric and of fainter maximum brightness. Its location in
the CMD suggests that it may be a foreground object. A similar
variable was found by Mochejska \& Kaluzny (1999) in NGC 7789 (V4),
displaying strong asymmetry of the light outside of the eclipses.

\begin{figure}[ht]
\plotfiddle{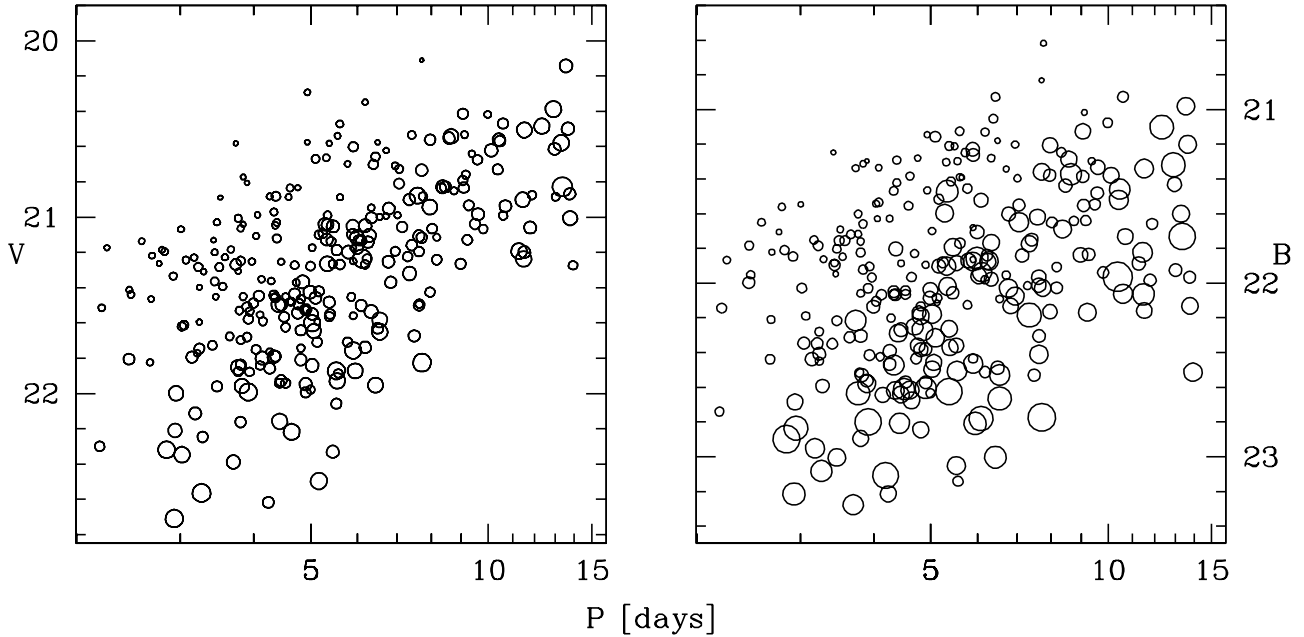}{8cm}{0}{90}{90}{-280}{-385}
\caption{The $V$ and $B$-band Period-Luminosity diagrams for the 
Cepheids in field M33B. The circles are proportional in size to 
their amplitudes in the corresponding band.}
\label{fig:pl}
\end{figure}

Most Cepheids occupy the region between $0.4<B-V<1$, with several
outliers stretching from $B-V=0$ to $B-V=1.2$. The discrepant colors
of some of the Cepheid variables are most likely caused by blending
with nearby bright stars of very different colors and/or by
reddening. The phenomenon of blending occurs when the Cepheid
possesses one or more close companions which cannot be separated at
the resolving power of the instrument used. A further discussion of
blending and its properties can be found in Mochejska et al. (2000;
2001) and Stanek \& Udalski (1999).

The other periodic variables are located all over the CMD. Most of
them seem to be pulsating variables. One of them,
D33J013359.8+303354.9, exhibits variability of an unclear nature. It
is located on the upper main sequence, with $V=17.89$ and
$B-V=-0.08$. Its light curve (Fig. \ref{fig:per}) shows two maxima of
unequal brightness with sharp minima in between, reminiscent of 
eclipses. This behavior is confirmed in our earlier data (Paper VI). 

The miscellaneous variables are also spread throughout the CMD, with
concentrations on the upper main sequence and the upper red giant
branch (RGB). Very few of the newly discovered variables are redder
than $B-V=1.5$. Most of the variables classified as long period are
located on the upper main sequence. Several exhibit colors similar to
Cepheids, but are much brighter. It is likely that some of these
variables are Cepheids with periods in excess of 14 days.

\begin{figure}[p]
\plotfiddle{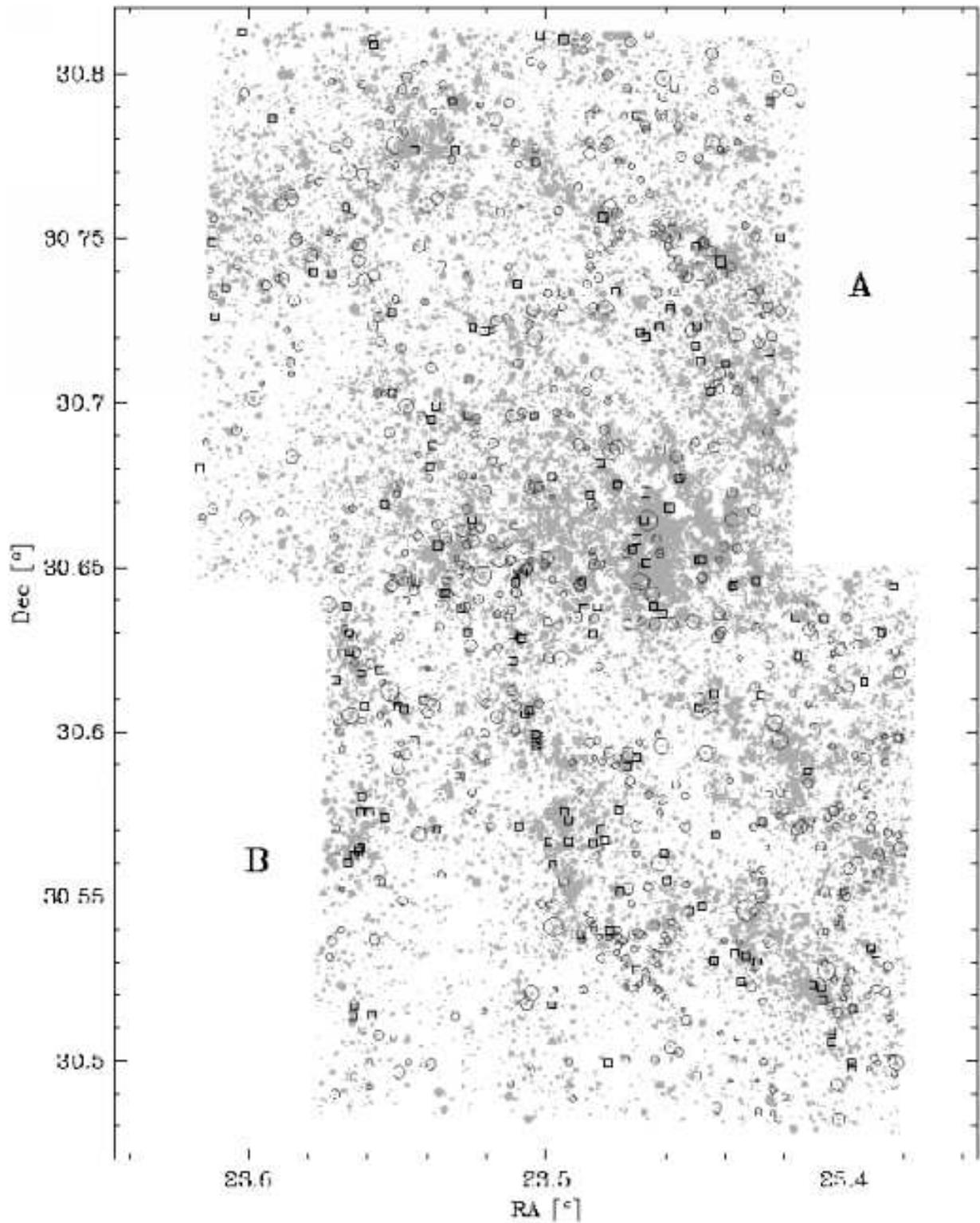}{20.5cm}{0}{100}{100}{-300}{-100}
\caption{The location of EBs (squares) and Cepheids (circles) in the
fields M33A (top) and M33B (bottom). The size of the Cepheid symbols
are proportional to their period. All other stars in the field with
$V<21.5$ mag are drawn as dots of size proportional to their
magnitude.}
\label{fig:xy}
\end{figure}

In Figure \ref{fig:pl} we present the $B$ and $V$-band P-L diagrams
for the Cepheid variables, drawn as open circles proportional in size
to their amplitudes. As expected, the amplitudes in the $B$-band are
on average larger than in $V$. A clear relation between the period and
magnitude is discernible. Several Cepheids too faint for their
periods are probably suffering from greater than average reddening.
Some of the faintest Cepheids exhibit quite large amplitudes. This is
not a physical effect and is most likely caused by a similar
phenomenon to the one discussed in Subsection \ref{sect:cmp} regarding
the comparison with Paper VI photometry. If the magnitude used to
convert the light curve from flux to magnitudes is measured too faint,
the resulting amplitude will be too large. 

For periods shorter than seven days the relations widen upwards
considerably, with the brighter Cepheids for a given period having
smaller amplitudes. One possibility is that these Cepheids are
pulsating in the first overtone. We have already found a population of
such stars in Paper VII, with eight very promising candidates among
them. On the other hand these could be fundamental mode Cepheids which
are heavily affected by blending: adding a constant flux would tend to
increase the brightness of a Cepheid and diminish its amplitude. We
will examine these stars in more depth in Section \ref{sect:fo}.

\begin{figure}[ht]
\plotfiddle{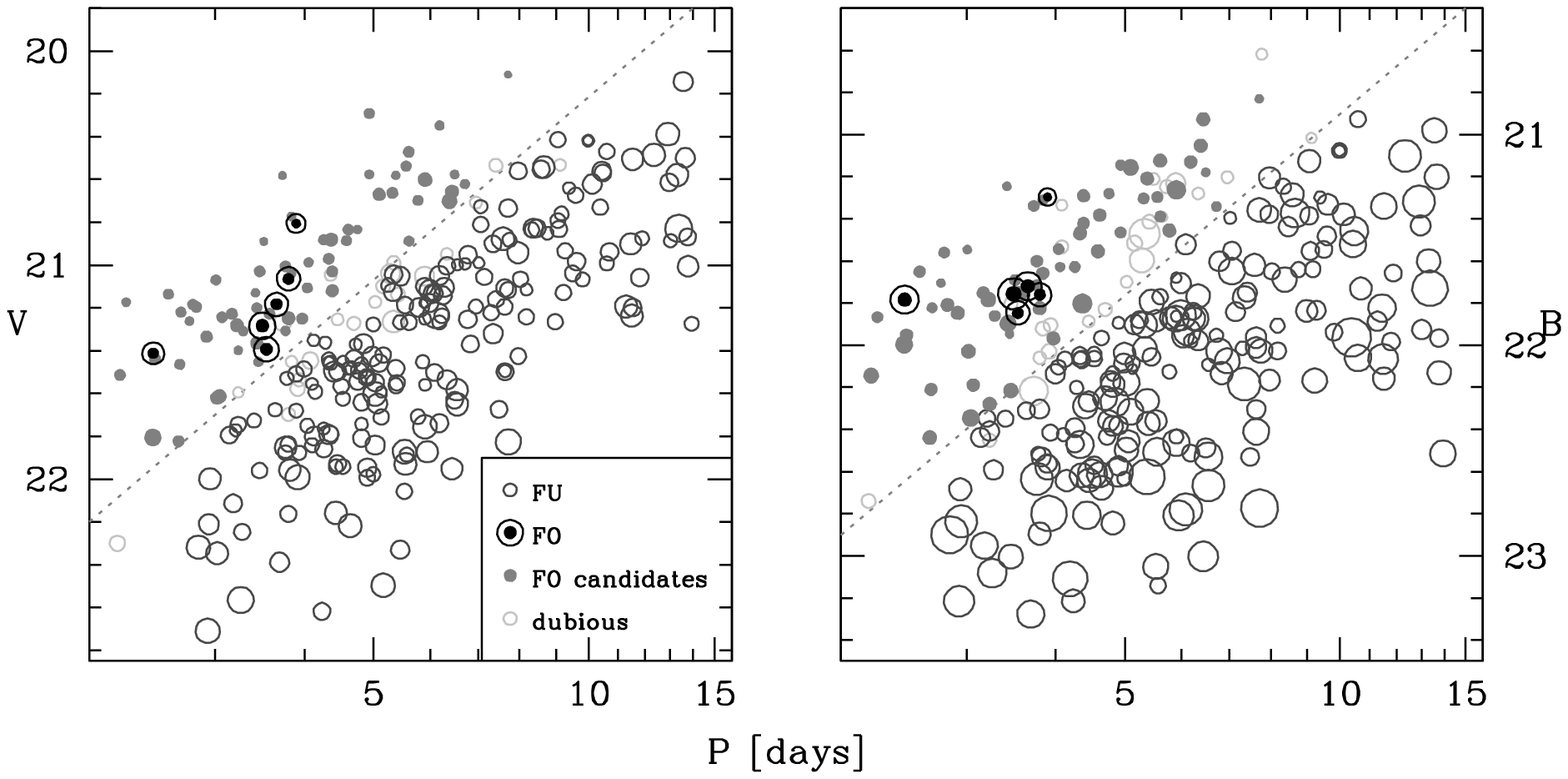}{8cm}{0}{90}{90}{-280}{-385}
\caption{The $V$ and $B$-band Period-Luminosity diagrams for the
Cepheids in field M33B. The circles are proportional in size to the
pulsational amplitudes of the Cepheids in the corresponding band. The
dotted lines show the division between FU (open circles) and FO
(filled circles) Cepheids. The encircled dots represent the most
reliable FO Cepheid candidates. The light open circles show the
Cepheids which are above the line in one band only.}
\label{fig:pl_fo}
\end{figure}

Figure \ref{fig:xy} shows the location of the EBs and Cepheids within
the field M33B. The Cepheids are plotted as open circles, proportional
in size to their period and the EBs as open squares. To trace the
spiral pattern of the galaxy we plot in the background all the stars
with $V<21.5$ mag as filled dots of size proportional to their
magnitude. These two types of variables appear somewhat more plentiful
within the spiral arms.

\section{First Overtone Cepheids}
\label{sect:fo}

As we have noted in the previous Section, on the $B$ and $V$-band P-L
diagrams in Fig. \ref{fig:pl} there are Cepheids which seem too bright
for their periods. In addition they possess smaller pulsational
amplitudes of variability compared to the normal Cepheids. Their
positions on the P-L diagrams would lead us to expect that these
should be first overtone (FO) pulsators (as in Fig. 2 of Udalski et
al. 1999; hereafter U99). The situation is, however, complicated by
the existence of blending. As a result of blending the Cepheid should
appear brighter because of the added constant flux and its amplitude,
measured in magnitudes, should decrease (Mochejska et al. 2000, 2001;
Stanek \& Udalski 1999).

\begin{figure}[ht]
\plotfiddle{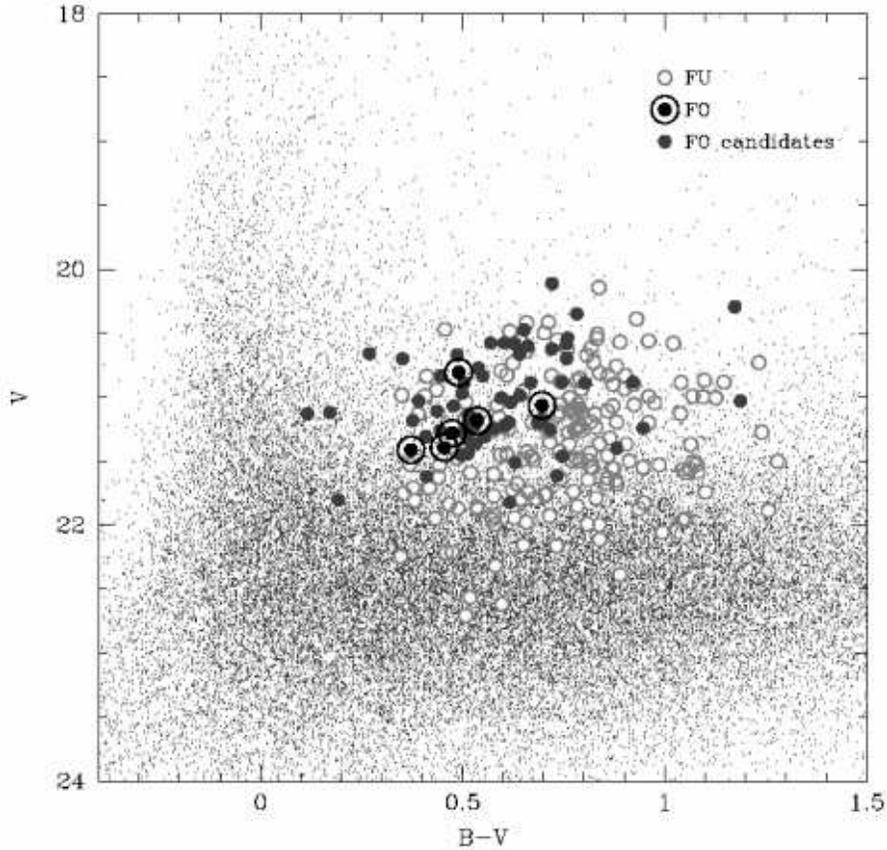}{10.5cm}{0}{100}{100}{-320}{-230}
\caption{The $V/B-V$ CMD for the Cepheid variables. The FU Cepheids are
denoted by open circles and the FO ones by filled circles. The encircled
dots represent the most reliable FO Cepheid candidates.}
\label{fig:cmd_fo}
\end{figure}

In order to try to determine whether these Cepheids are first overtone
pulsators, we have checked whether they possess other properties
expected of such stars. In addition to being brighter and having a
smaller amplitude, Cepheids pulsating in the first overtone should
have more symmetric (sinusoidal) light curves and be on average bluer
than fundamental mode Cepheids (FU). The most powerful technique for
discriminating them from FU Cepheids are the Fourier parameters of
their light curves (Antonello \& Aikawa 1995; Beaulieu et al. 1995).

To select a sample of FO Cepheid candidates we have made a division
on the P-L diagrams roughly parallel to the P-L relation, between the
bright low amplitude and fainter high amplitude Cepheids (dotted lines
on Fig. \ref{fig:pl_fo}). In our final sample we included the Cepheids
which were above these lines in both of the P-L diagrams. There are 44
such Cepheids in our catalog (filled circles). FO Cepheids should have
periods ranging from 1.7 to 6 days.  We decided not to make a cutoff
at higher periods, although we did not regard it likely that Cepheids
with $P>6$ days would turn out to be FO pulsators.

\begin{figure}[ht]
\plotfiddle{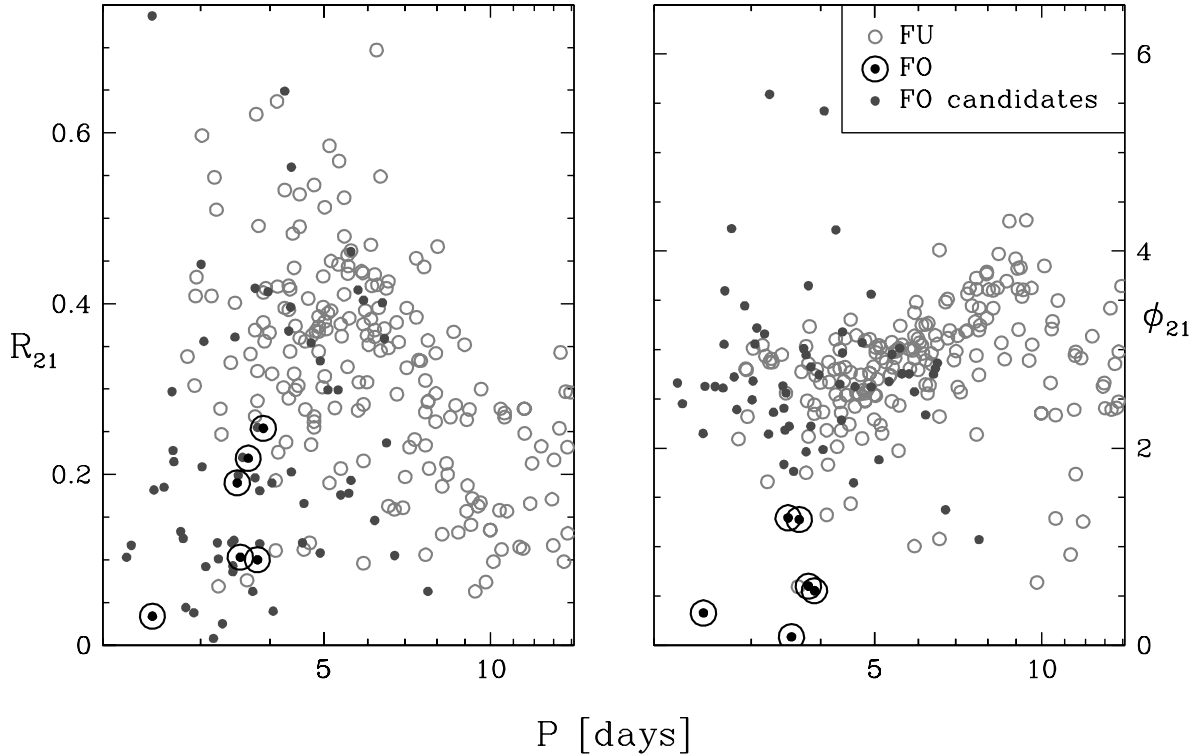}{9.2cm}{0}{80}{80}{-260}{-315}
\caption{The Fourier parameters $R_{21}$ and $\phi_{21}$ of the
Cepheid light curves as a function of period. The FU Cepheids are
denoted by open circles and the FO ones by filled circles. The
encircled dots represent the most reliable FO Cepheid candidates.}
\label{fig:pars}
\end{figure}

On the $V/B-V$ CMD (Fig. \ref{fig:cmd_fo}) we plot the positions of
the FU Cepheids (open circles) and FO candidates (filled circles). As
expected of FO Cepheids, our candidates are bluer than the FU ones. We
note that on the CMD presented by U99 for the Large Magellanic Cloud
(LMC) Cepheids (Fig. 4 therein) there is also some overlap between the
loci of those two types of Cepheids.

The relations between the light curve Fourier parameters
$R_{21}=A_2/A_1$, $\phi_{21}=\phi_2-2\phi_1$ and the period for the FO
Cepheid candidates (filled circles) and FU mode pulsators (open
circles) are shown in Fig. \ref{fig:pars}. We have compared them with
Fig. 3 in U99 for LMC Cepheids. We notice on the $R_{21}/\log P$
diagram for LMC that within our range of periods the FO Cepheids
progress upwards in $R_{21}$ with increasing period, forming the
second branch of the V-shaped pattern and then merge with the FU
Cepheid sequence at higher periods. The situation is very similar on
the $\phi_{21}/\log P$ diagram in U99, where on the one hand the FU
Cepheids are confined to a narrower sequence, but on the other there
is more overlap between them and the FO pulsators.

\begin{figure}[ht]
\plotfiddle{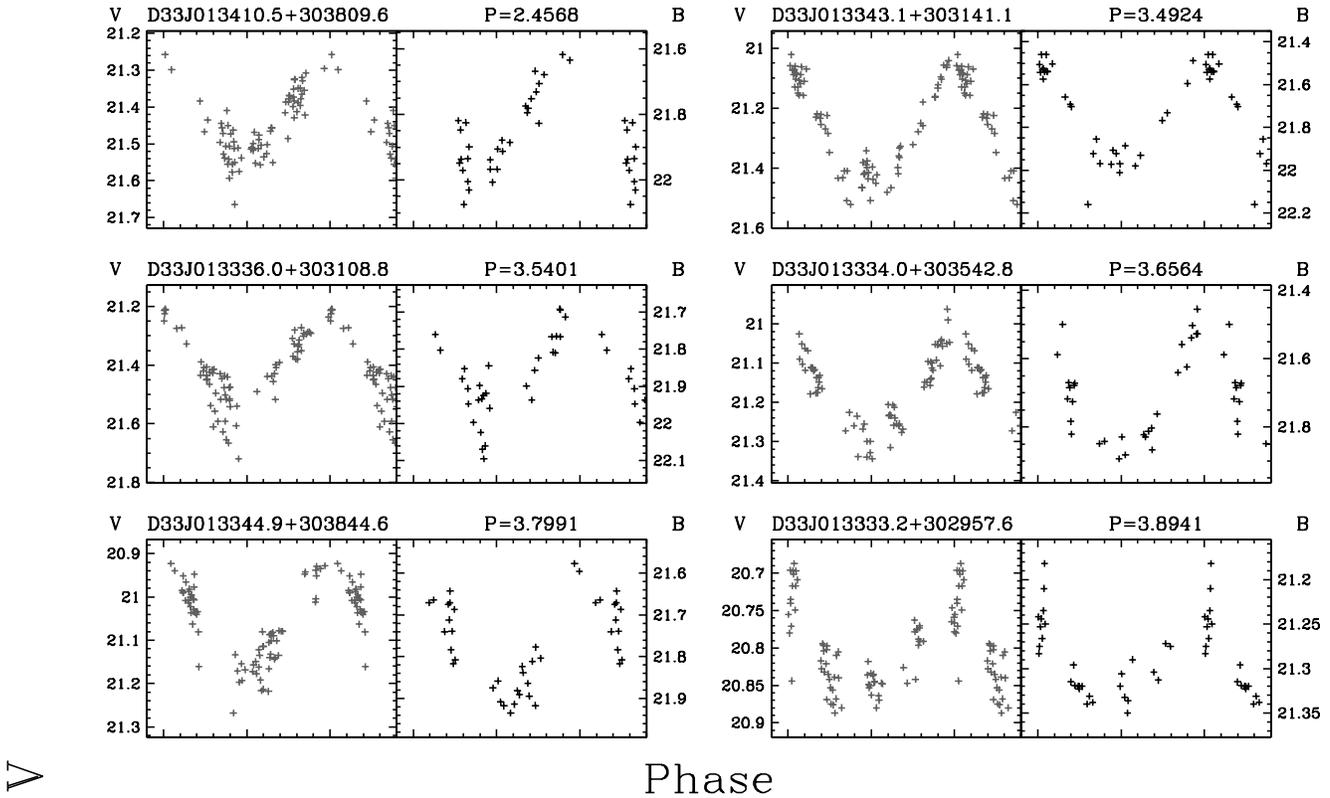}{9.5cm}{0}{88}{88}{-280}{-360}
\caption{The $V$ and $B$-band light curves of the six FO Cepheids
in the field M33B.}
\label{fig:lc_fo}
\end{figure}

Indeed we do observe in Fig. \ref{fig:pars} the FO Cepheid candidates
to occupy roughly the predicted positions despite our larger scatter
in the Fourier parameters than in U99. From the $\phi_{21}/\log P$
diagram in the right panel we have selected six FO Cepheids which are
most separated from the FU Cepheid sequence. We examined their
positions on the $R_{21}$ diagram to find that they are also rather
well separated from the FU mode Cepheids. These six bona fide FO
Cepheid candidates are denoted by encircled dots on Figs
\ref{fig:pl_fo}-\ref{fig:pars}. As for the rest of the candidates, we
believe many of them may also be FO Cepheids. More accurate photometry
obtained with a larger instrument would be necessary to obtain better
Fourier parameters of their light curves and resolve this issue.

In Figure \ref{fig:lc_fo} we show the light curves of the six FO
Cepheids. They appear more symmetrical in shape and do not exhibit the
fast rise and slow decline typical for FU Cepheids (see Fig.
\ref{fig:ceph}). This fact lends further credibility to the notion 
that these Cepheids pulsate in the first overtone.

\section{Conclusions}
Our search for variable stars in the data from the followup
observations of the detached eclipsing binary D33J013337.0+303032.8 in
field M33B collected at the 2.1m KPNO telescope resulted in the
discovery of 96 eclipsing binaries, 349 Cepheids, and 450 other
periodic, possible long period or non-periodic variables. Out of the
total 895 variables 612 are not listed in our other M33 catalogs
(Papers VI and VII). Due to the short time base of our observations,
we were limited to variables with periods not exceeding 14
days. Thanks to the use of a larger aperture instrument and a novel
method of image reduction -- the ISIS image subtraction package, we
were more efficient at finding the fainter and lower amplitude
variables than in our previous study of this field, especially for
short period Cepheids ($P<8d$). We have also found a population of
Cepheids which are most likely pulsating in the first overtone and for
six of them we present strong arguments in favor of this
interpretation.

The method of image subtraction has two main advantages over the
classical profile fitting method. It is more efficient in discovering
variables: in field M33A we have discovered 355 periodic variables
using ISIS and only 212 with Dophot (Paper VII). Additionally in
crowded fields image subtraction can lead to large improvements in the
photometric accuracy (Alard 2000b).

\acknowledgments{We thank the TAC of the KPNO for the generous
allocation of the observing time. We would like to thank Lucas Macri
for supplying us with the FLWO data for two variables, Grzegorz
Pojma{\'n}ski for $lc$ - the light curve analysis utility and Wojtek
Pych and Alex Schwarzenberg-Czerny for their software. BJM was
supported by the Polish KBN grant 2P03D025.19 and the Foundation for
Polish Science stipend for young scientists. JK was supported by
the NSF grant AST-9819787. DDS acknowledges support from the Alfred
P. Sloan Foundation and from NSF grant No. AST-9970812.}

\begin{small}
\tablenum{1}
\begin{planotable}{lrrrrrcrl}
\tablewidth{0pc}
\tablecaption{\sc DIRECT Eclipsing Binaries in M33B}
\tablehead{
\colhead{} & \colhead{$P$} & \colhead{} & \colhead{} & \colhead{} &
\colhead{} & \colhead{$i$} & \colhead{} & \colhead{} \\
\colhead{Name} & \colhead{$(days)$} & \colhead{$V_{max}$} & \colhead{$B_{max}$} &
\colhead{$R_1$} & \colhead{$R_2$} & \colhead{(deg)} & \colhead{$e$} &
\colhead{Comments}}
\startdata
 D33J013359.5+303334.7 &  0.8682  & \nodata &   20.60 & 0.57 & 0.39 & 55.83 & 0.01 &     \nl 
 D33J013359.5+303101.7 &  0.9162  &   18.05 &   19.11 & 0.30 & 0.23 & 68.90 & 0.03 &     \nl 
 D33J013350.5+303346.8 &  1.0757  &   20.43 &   20.43 & 0.62 & 0.37 & 54.02 & 0.03 &     \nl 
 D33J013356.2+303748.1 &  1.1047: & \nodata &   21.97 & 0.44 & 0.44 & 76.64 & 0.30 &     \nl 
 D33J013357.0+303814.4 &  1.1078  & \nodata &   21.82 & 0.56 & 0.44 & 83.94 & 0.00 &     \nl 
 D33J013415.8+303727.4 &  1.2019  &   20.71 & \nodata & 0.49 & 0.30 & 66.69 & 0.00 &     \nl 
 D33J013401.9+303741.8 &  1.2597  &   20.13 &   20.15 & 0.48 & 0.33 & 73.70 & 0.09 &     \nl 
 D33J013414.6+303628.2 &  1.3234  &   21.15 &   21.22 & 0.44 & 0.37 & 85.55 & 0.01 &     \nl 
 D33J013406.8+303816.2 &  1.3456  &   20.35 &   20.49 & 0.57 & 0.34 & 51.31 & 0.06 &     \nl 
 D33J013415.0+303352.6 &  1.6468  & \nodata &   19.73 & 0.63 & 0.36 & 51.03 & 0.00 &     \nl 
 D33J013414.0+303049.8 &  1.6681: &   20.35 &   20.99 & 0.67 & 0.33 & 46.42 & 0.04 &     \nl 
 D33J013332.0+303839.2 &  1.6691  &   19.70 &   19.72 & 0.56 & 0.37 & 65.12 & 0.00 & 1   \nl 
 D33J013416.1+303817.2 &  1.6818  &   20.39 &   20.64 & 0.34 & 0.24 & 67.60 & 0.01 &     \nl 
 D33J013354.2+303435.2 &  1.7433: &   20.57 &   20.53 & 0.35 & 0.27 & 74.67 & 0.01 &     \nl 
 D33J013410.4+303843.7 &  1.8115  &   21.83 &   21.76 & 0.36 & 0.27 & 77.67 & 0.01 &     \nl 
 D33J013415.0+303431.5 &  1.8763  &   19.85 &   19.82 & 0.59 & 0.33 & 60.48 & 0.00 & 1   \nl 
 D33J013347.4+303249.0 &  1.8990  &   20.58 &   20.69 & 0.46 & 0.27 & 71.28 & 0.02 & 1   \nl 
 D33J013337.6+303803.1 &  1.9020  &   21.02 &   20.90 & 0.47 & 0.40 & 85.85 & 0.02 & 1   \nl 
 D33J013400.9+303556.7 &  2.0806  &   20.73 &   20.74 & 0.46 & 0.31 & 72.48 & 0.01 &     \nl 
 D33J013350.2+303317.8 &  2.0991  &   20.16 &   20.42 & 0.46 & 0.38 & 85.27 & 0.05 & 1   \nl 
 D33J013342.6+303420.8 &  2.1037  &   20.26 &   20.24 & 0.65 & 0.34 & 49.93 & 0.00 &     \nl 
 D33J013408.8+303412.5 &  2.2115  &   20.11 &   19.98 & 0.40 & 0.26 & 76.15 & 0.00 &     \nl 
 D33J013415.8+303748.5 &  2.3258  &   21.46 &   21.36 & 0.47 & 0.32 & 85.75 & 0.01 &     \nl 
 D33J013415.1+303349.6 &  2.3288  & \nodata &   19.74 & 0.47 & 0.35 & 70.66 & 0.01 &     \nl 
 D33J013356.2+303358.6 &  2.3362  &   19.10 &   18.94 & 0.59 & 0.40 & 80.22 & 0.00 & 1   \nl 
 D33J013402.2+303416.8 &  2.3697  &   19.99 &   19.77 & 0.67 & 0.33 & 53.02 & 0.00 &     \nl 
 D33J013347.6+303625.8 &  2.3708  &   20.74 &   20.94 & 0.38 & 0.21 & 75.21 & 0.02 &     \nl 
 D33J013402.7+303746.6 &  2.4162  &   20.74 &   21.01 & 0.25 & 0.18 & 76.76 & 0.00 &     \nl 
 D33J013355.8+303816.7 &  2.4463  &   21.63 &   21.68 & 0.35 & 0.28 & 84.76 & 0.02 &     \nl 
 D33J013346.5+303642.0 &  2.5123  &   19.90 &   19.89 & 0.36 & 0.22 & 65.50 & 0.01 &     \nl 
 D33J013339.8+303806.0 &  2.5165  &   19.14 &   19.08 & 0.62 & 0.38 & 46.32 & 0.00 & 1   \nl 
 D33J013334.4+303654.9 &  2.5272  &   21.41 &   21.25 & 0.23 & 0.15 & 81.08 & 0.06 &     \nl 
 D33J013408.2+303831.8 &  2.7174  &   21.34 &   21.10 & 0.44 & 0.39 & 88.66 & 0.03 &     \nl 
 D33J013358.6+303431.6 &  2.7335  &   19.91 &   19.76 & 0.62 & 0.38 & 47.47 & 0.01 &     \nl 
 D33J013407.0+303509.6 &  2.7820: &   21.25 &   21.16 & 0.75 & 0.09 & 51.93 & 0.23 &     \nl 
 D33J013353.4+303522.6 &  2.7995  &   20.49 &   20.47 & 0.40 & 0.26 & 72.40 & 0.02 &     \nl 
 D33J013359.8+303359.7 &  2.8392  &   20.76 &   20.58 & 0.40 & 0.33 & 80.08 & 0.07 &     \nl 
 D33J013338.9+303516.4 &  2.8464  &   21.08 &   20.79 & 0.43 & 0.38 & 80.26 & 0.03 &     \nl 
 D33J013355.3+303401.8 &  2.8856  &   22.49 &   22.13 & 0.61 & 0.38 & 88.63 & 0.00 &     \nl 
 D33J013414.3+303432.5 &  2.8878  &   20.81 & \nodata & 0.46 & 0.28 & 62.85 & 0.05 &     \nl 
 D33J013352.0+303905.0 &  2.9592  & \nodata &   19.12 & 0.59 & 0.40 & 86.71 & 0.01 & 1   \nl 
 D33J013335.4+302952.3 &  2.9756  & \nodata &   19.97 & 0.37 & 0.26 & 66.84 & 0.10 &     \nl 
 D33J013333.8+303203.1 &  3.0504  &   20.68 & \nodata & 0.60 & 0.40 & 49.83 & 0.02 &     \nl 
 D33J013342.6+303640.3 &  3.0562  &   21.45 &   22.04 & 0.42 & 0.28 & 63.56 & 0.03 &     \nl 
 D33J013350.7+303808.4 &  3.0632  &   19.89 &   19.81 & 0.63 & 0.36 & 61.79 & 0.00 &     \nl 
 D33J013337.9+303120.7 &  3.0678  &   19.55 &   19.48 & 0.57 & 0.42 & 61.66 & 0.03 & 1   \nl 
 D33J013344.7+303157.4 &  3.0931  &   21.27 &   21.23 & 0.37 & 0.17 & 68.61 & 0.16 &     \nl 
 D33J013351.4+303817.0 &  3.0968: & \nodata &   19.84 & 0.35 & 0.27 & 89.71 & 0.08 &     \nl 
 D33J013346.4+303407.6 &  3.1525  &   19.56 &   19.44 & 0.61 & 0.36 & 72.76 & 0.03 & 1   \nl 
 D33J013337.7+303106.2 &  3.1855  & \nodata &   18.75 & 0.63 & 0.22 & 40.42 & 0.16 &     \nl 
 D33J013335.3+302958.0 &  3.2232  &   20.31 &   20.17 & 0.34 & 0.24 & 68.77 & 0.17 &     \nl 
 D33J013413.0+303426.7 &  3.2633  &   20.17 &   20.07 & 0.32 & 0.23 & 82.42 & 0.01 & 1   \nl 
 D33J013344.3+303127.0 &  3.3001  &   20.72 &   20.58 & 0.55 & 0.44 & 89.06 & 0.07 &     \nl 
 D33J013343.9+303155.1 &  3.3293  &   19.17 &   19.02 & 0.69 & 0.31 & 50.69 & 0.07 &     \nl 
 D33J013415.5+303344.7 &  3.3570  & \nodata &   19.61 & 0.55 & 0.33 & 52.34 & 0.00 &     \nl 
 D33J013335.3+303057.5 &  3.4769  &   20.96 &   21.21 & 0.63 & 0.34 & 45.80 & 0.01 &     \nl 
 D33J013415.7+303047.8 &  3.5157: &   19.26 &   19.27 & 0.71 & 0.28 & 50.19 & 0.01 &     \nl 
 D33J013339.7+303722.1 &  3.5808  &   22.16 &   22.03 & 0.31 & 0.26 & 85.77 & 0.00 &     \nl 
 D33J013417.0+303655.9 &  3.6574  &   19.57 &   19.42 & 0.38 & 0.21 & 74.77 & 0.01 &     \nl 
 D33J013413.3+303707.7 &  3.6863  &   19.89 &   20.09 & 0.53 & 0.33 & 57.40 & 0.01 &     \nl 
 D33J013346.5+303150.0 &  3.7427  &   21.92 &   21.91 & 0.44 & 0.40 & 80.84 & 0.04 &     \nl 
 D33J013400.9+303545.1 &  3.8955  &   20.16 &   20.12 & 0.47 & 0.47 & 72.50 & 0.01 & 1   \nl 
 D33J013358.1+303400.2 &  4.0523  &   19.66 &   19.58 & 0.63 & 0.35 & 43.29 & 0.00 & 1   \nl 
 D33J013415.4+303100.6 &  4.5675  &   18.91 &   18.83 & 0.25 & 0.14 & 75.57 & 0.02 &     \nl 
 D33J013331.6+303554.2 &  4.7624  &   21.21 &   21.26 & 0.43 & 0.31 & 72.88 & 0.03 &     \nl 
 D33J013357.2+303219.0 &  4.9380  &   19.11 &   18.98 & 0.51 & 0.49 & 47.80 & 0.01 & 1   \nl 
 D33J013355.1+302957.4 &  5.0920  &   20.04 &   19.92 & 0.36 & 0.22 & 77.50 & 0.08 & 1   \nl 
 D33J013352.6+303138.7 &  5.1492  &   21.17 &   21.50 & 0.67 & 0.33 & 58.36 & 0.03 &     \nl 
 D33J013411.9+303628.0 &  5.2239  &   20.25 &   20.31 & 0.39 & 0.29 & 72.97 & 0.02 & 1   \nl 
 D33J013358.1+303424.3 &  5.3011  &   19.64 &   19.52 & 0.57 & 0.43 & 44.09 & 0.03 &     \nl 
 D33J013342.5+303314.4 &  5.3325  & \nodata &   18.79 & 0.30 & 0.23 & 66.81 & 0.09 & 1   \nl 
 D33J013338.5+303124.2 &  5.5371  &   18.90 &   18.80 & 0.50 & 0.45 & 50.32 & 0.03 & 1   \nl 
 D33J013352.7+303532.3 &  5.5966  & \nodata &   19.98 & 0.54 & 0.41 & 67.15 & 0.02 & 1   \nl 
 D33J013348.5+303243.7 &  6.0990  &   21.34 &   21.64 & 0.29 & 0.28 & 79.84 & 0.17 &     \nl 
 D33J013337.0+303032.8 &  6.1626  &   19.60 &   19.44 & 0.19 & 0.14 & 79.50 & 0.23 & 1   \nl 
 D33J013401.7+303619.9 &  6.2788: & \nodata &   19.10 & 0.63 & 0.26 & 44.99 & 0.06 &     \nl 
 D33J013410.7+303550.1 &  6.3837  &   20.30 &   20.35 & 0.59 & 0.40 & 56.66 & 0.01 &     \nl 
 D33J013402.7+303718.2 &  6.4280  &   20.46 &   21.07 & 0.30 & 0.30 & 71.48 & 0.22 &     \nl 
 D33J013343.0+303149.1 &  6.4993  &   20.22 &   20.20 & 0.61 & 0.39 & 52.56 & 0.04 &     \nl 
 D33J013333.4+303159.2 &  6.5015  &   21.12 &   21.57 & 0.60 & 0.40 & 58.45 & 0.00 &     \nl 
 D33J013354.8+303222.6 &  6.6518: &   18.22 &   18.28 & 0.71 & 0.29 & 51.12 & 0.08 &     \nl 
 D33J013333.0+303749.8 &  6.6860  &   20.04 &   20.01 & 0.38 & 0.25 & 67.31 & 0.07 &     \nl 
 D33J013406.4+303749.6 &  7.0058  &   18.82 &   18.79 & 0.29 & 0.19 & 65.50 & 0.05 &     \nl 
 D33J013344.8+303839.4 &  7.1830  & \nodata &   19.81 & 0.44 & 0.39 & 64.07 & 0.02 &     \nl 
 D33J013355.6+303412.8 &  7.4391  &   19.67 &   19.54 & 0.24 & 0.18 & 78.19 & 0.15 & 1   \nl 
 D33J013354.0+303304.5 &  8.7761  & \nodata &   19.03 & 0.48 & 0.36 & 72.21 & 0.01 & 1   \nl 
 D33J013336.9+303019.8 &  9.0223  &   19.21 &   19.19 & 0.67 & 0.33 & 60.10 & 0.07 & 1   \nl 
 D33J013409.8+303634.6 &  9.2676  &   21.39 &   21.36 & 0.27 & 0.22 & 89.09 & 0.01 &     \nl 
 D33J013411.5+303625.4 &  9.8310  &   18.32 &   18.44 & 0.68 & 0.31 & 45.79 & 0.03 &     \nl 
\enddata
\tablecomments{(1) Variables identified by Macri et al. (2001a)}
\label{tab:ecl}
\end{planotable}
\end{small}

\begin{small}
\tablenum{2}
\begin{planotable}{lll}
\tablewidth{0pc}
\tablecaption{\sc DIRECT Flux Eclipsing Binaries in M33B}
\tablehead{
\colhead{} & \colhead{$P$} & \colhead{}\\
\colhead{Name} & \colhead{$(days)$} & \colhead{Comments}}
\startdata
 D33J013414.9+303449.7 &  1.6767 &      \nl 
 D33J013357.2+303846.1 &  2.2157 &      \nl 
 D33J013402.4+303842.4 &  2.4338 &      \nl 
 D33J013401.4+303623.5 &  2.5535 &      \nl 
 D33J013343.1+303845.1 &  2.8873 & 1    \nl 
 D33J013414.9+303703.6 &  6.1083 &      \nl 
 D33J013416.0+303337.5 &  7.0801 &      \nl 
\enddata   
\tablecomments{(1) Variables identified by Macri et al. (2001a)}
\label{tab:ecl_flux}
\end{planotable}
\end{small}

\begin{small}
\tablenum{3}
\begin{planotable}{ccccc}
\tablewidth{0pc}
\tablecaption{\sc Light Curves of Eclipsing Binaries in M33B}
\tablehead{ \colhead{Name} & \colhead{Filter} &
\colhead{HJD$-$2451000} & \colhead{mag} & \colhead{$\sigma_{mag}$} }
\startdata
D33J013408.2+303831.8 & B & 452.7982 &     21.353 &   0.027 \\
                      & B & 452.9166 &     21.214 &   0.033 \\
                      & B & 454.6968 &     21.172 &   0.025 \\
                      & B & 454.7299 &     21.153 &   0.019 \\
\enddata
\tablecomments{Table 3 is available in its entirety in the electronic
version of the Astronomical Journal. A portion is shown here for
guidance regarding its form and content.}
\label{tab:lc_ecl}
\end{planotable}
\end{small}
 
\begin{small}
\tablenum{4}
\begin{planotable}{lrrrrrl}
\tablewidth{0pc}
\tablecaption{\sc DIRECT Cepheids in M33B}
\tablehead{
\colhead{} & \colhead{$P$} & \colhead{} & \colhead{} & \colhead{} &
\colhead{} & \colhead{} \\
\colhead{Name} & \colhead{$(days)$} & \colhead{$\langle V\rangle$} &
\colhead{$\langle B\rangle$} & \colhead{$A_V$} & \colhead{$A_B$} &
\colhead{Comments}}
\startdata
 D33J013414.2+303000.0 &  2.188: &   22.30 &   22.74 &    0.23 &    0.20 &     \nl 
 D33J013348.4+303124.1 &  2.206  &   21.51 &   22.14 &    0.14 &    0.23 &     \nl 
 D33J013334.3+303448.0 &  2.251: &   21.17 &   21.87 &    0.09 &    0.15 &     \nl 
 D33J013412.4+303850.9 &  2.454  &   21.80 &   22.00 &    0.27 &    0.30 &     \nl 
 D33J013410.5+303809.6 &  2.457  &   21.41 &   21.78 &    0.12 &    0.20 &     \nl 
 D33J013353.7+303143.5 &  2.472  &   21.44 &   21.95 &    0.14 &    0.18 &     \nl 
 D33J013353.6+303120.9 &  2.579  &   21.14 &   21.65 &    0.11 &    0.16 &     \nl 
 D33J013344.7+303024.6 &  2.665: &   21.82 &   22.44 &    0.13 &    0.22 &     \nl 
 D33J013356.7+303241.1 &  2.676  &   21.46 &   22.21 &    0.11 &    0.17 &     \nl 
 D33J013343.0+302938.5 &  2.686  &   21.22 &   21.82 &    0.11 &    0.11 &     \nl 
 D33J013345.2+302938.5 &  2.762  &   21.26 &   21.70 &    0.08 &    0.09 &     \nl 
 D33J013416.5+303307.2 &  2.791  &   21.18 &   21.56 &    0.11 &    0.14 &     \nl 
 D33J013405.6+303536.8 &  2.822  &   21.20 &   21.81 &    0.14 &    0.21 &     \nl 
 D33J013336.9+303725.8 &  2.842  &   22.32 &   22.90 &    0.46 &    0.81 &     \nl 
 D33J013406.0+303857.9 &  2.917  &   21.33 &   21.85 &    0.16 &    0.21 &     \nl 
 D33J013356.1+303531.2 &  2.928: &   22.71 &   23.21 &    0.49 &    0.64 &     \nl 
 D33J013416.8+303347.9 &  2.939  &   22.21 &   22.68 &    0.36 &    0.43 &     \nl 
 D33J013409.6+303709.6 &  2.950  &   22.00 &   22.84 &    0.39 &    0.69 &     \nl 
 D33J013348.9+303056.0 &  3.007: &   21.07 &   21.55 &    0.12 &    0.09 &     \nl 
 D33J013332.7+303555.0 &  3.019: &   21.62 &   22.03 &    0.19 &    0.22 &     \nl 
 D33J013342.2+303353.1 &  3.019: &   22.35 & \nodata &    0.43 & \nodata &     \nl 
 D33J013351.0+303557.7 &  3.042  &   21.61 &   22.35 &    0.20 &    0.30 &     \nl 
 D33J013336.6+303230.0 &  3.066  &   21.24 &   22.19 &    0.14 &    0.18 &     \nl 
 D33J013345.9+303030.3 &  3.140: &   21.79 &   22.44 &    0.29 &    0.35 &     \nl 
 D33J013412.4+303739.3 &  3.166  &   21.23 &   21.75 &    0.13 &    0.18 &     \nl 
 D33J013359.4+303117.4 &  3.179: &   22.11 &   22.95 &    0.31 &    0.54 &     \nl 
 D33J013352.8+303346.1 &  3.205: &   21.77 &   22.35 &    0.14 &    0.29 &     \nl 
 D33J013412.8+303607.4 &  3.218  &   21.28 &   21.78 &    0.19 &    0.23 &     \nl 
 D33J013340.0+303404.8 &  3.232: &   21.59 &   22.45 &    0.10 &    0.18 &     \nl 
 D33J013343.3+303323.5 &  3.232  &   21.40 &   22.28 &    0.08 &    0.19 &     \nl 
 D33J013407.0+303850.1 &  3.232  &   21.74 &   22.41 &    0.25 &    0.33 &     \nl 
 D33J013343.3+303631.0 &  3.259  &   22.56 &   23.08 &    0.51 &    0.60 &     \nl 
 D33J013331.9+302945.6 &  3.272  &   22.25 &   22.59 &    0.26 &    0.35 &     \nl 
 D33J013414.2+302931.2 &  3.286: &   21.31 &   21.86 &    0.10 &    0.13 &     \nl 
 D33J013345.6+303211.6 &  3.401: &   21.73 &   22.35 &    0.20 &    0.26 &     \nl 
 D33J013344.5+303209.8 &  3.416: &   21.13 &   21.25 &    0.08 &    0.07 &     \nl 
 D33J013402.9+302934.5 &  3.431  &   21.20 &   21.88 &    0.12 &    0.16 &     \nl 
 D33J013405.9+303433.0 &  3.431  &   21.36 &   21.90 &    0.18 &    0.27 &     \nl 
 D33J013334.1+303111.4 &  3.446  &   21.45 &   21.95 &    0.10 &    0.09 &     \nl 
 D33J013336.4+303258.9 &  3.461: & \nodata &   21.80 & \nodata &    0.18 &     \nl 
 D33J013344.4+303720.6 &  3.461  &   21.03 &   22.22 &    0.13 &    0.23 &     \nl 
 D33J013401.1+302950.4 &  3.461: &   21.96 &   23.00 &    0.25 &    0.48 &     \nl 
 D33J013343.1+303141.1 &  3.492  &   21.28 &   21.76 &    0.18 &    0.23 &     \nl 
 D33J013333.9+302951.0 &  3.508: &   20.89 &   21.69 &    0.06 &    0.13 &     \nl 
 D33J013336.0+303108.8 &  3.540  &   21.39 &   21.85 &    0.16 &    0.14 &     \nl 
 D33J013404.7+303527.4 &  3.572  &   21.23 &   21.75 &    0.13 &    0.23 &     \nl 
 D33J013352.5+303124.3 &  3.639  &   21.68 &   22.31 &    0.17 &    0.27 &     \nl 
 D33J013334.0+303542.8 &  3.656  &   21.18 &   21.72 &    0.13 &    0.19 &     \nl 
 D33J013416.5+303540.4 &  3.691  &   22.39 &   23.28 &    0.35 &    0.57 &     \nl 
 D33J013352.9+303119.1 &  3.726  &   20.58 &   21.34 &    0.06 &    0.14 &     \nl 
 D33J013413.4+303618.1 &  3.726  &   21.27 &   22.22 &    0.27 &    0.59 &     \nl 
 D33J013350.9+303431.9 &  3.762  &   21.01 &   21.60 &    0.15 &    0.16 &     \nl 
 D33J013355.7+303214.9 &  3.762: &   21.31 &   21.72 &    0.10 &    0.14 &     \nl 
 D33J013407.1+303512.4 &  3.762  &   21.85 &   22.64 &    0.40 &    0.67 &     \nl 
 D33J013412.0+303535.0 &  3.762  &   21.88 & \nodata &    0.22 & \nodata &     \nl 
 D33J013412.3+303024.4 &  3.781  &   21.53 &   22.52 &    0.17 &    0.20 &     \nl 
 D33J013344.9+303844.6 &  3.799  &   21.06 &   21.76 &    0.13 &    0.15 &     \nl 
 D33J013352.5+303548.0 &  3.799: &   21.70 &   22.06 &    0.18 &    0.16 &     \nl 
 D33J013405.0+303054.3 &  3.799  &   21.25 &   21.82 &    0.20 &    0.20 &     \nl 
 D33J013406.5+303901.8 &  3.799  &   21.84 &   22.30 &    0.24 &    0.33 &     \nl 
 D33J013411.1+302958.1 &  3.799: &   22.16 &   22.89 &    0.26 &    0.43 &     \nl 
 D33J013350.4+303108.2 &  3.818: &   21.96 &   22.53 &    0.39 &    0.35 &     \nl 
 D33J013341.2+303428.1 &  3.837  &   20.77 &   21.31 &    0.09 &    0.12 &     \nl 
 D33J013341.7+302902.5 &  3.837  &   21.04 &   21.66 &    0.09 &    0.18 &     \nl 
 D33J013406.9+303903.5 &  3.837  &   21.45 &   21.92 &    0.15 &    0.21 &     \nl 
 D33J013333.2+302957.6 &  3.894: &   20.81 &   21.30 &    0.07 &    0.05 &     \nl 
 D33J013335.3+303427.1 &  3.894  &   21.68 &   22.56 &    0.20 &    0.32 &     \nl 
 D33J013400.5+303550.2 &  3.894: &   21.51 &   22.58 &    0.23 &    0.49 &     \nl 
 D33J013359.7+303251.1 &  3.914: &   21.58 &   22.03 &    0.22 &    0.22 &     \nl 
 D33J013415.6+303744.3 &  3.914  &   21.99 &   22.80 &    0.48 &    0.78 &     \nl 
 D33J013411.7+303547.1 &  3.933  &   21.88 &   22.41 &    0.20 &    0.26 &     \nl 
 D33J013416.9+303610.9 &  3.933  &   21.54 &   21.90 &    0.18 &    0.19 &     \nl 
 D33J013341.2+302856.4 &  3.970  &   21.25 &   21.97 &    0.13 &    0.20 & 1   \nl 
 D33J013337.0+303544.2 &  3.994  &   21.49 &   22.14 &    0.23 &    0.33 &     \nl 
 D33J013338.8+303630.1 &  4.035  &   21.75 &   22.10 &    0.21 &    0.22 &     \nl 
 D33J013411.7+303843.5 &  4.035  &   21.11 &   21.54 &    0.12 &    0.14 &     \nl 
 D33J013357.9+302918.8 &  4.056  &   20.99 &   21.63 &    0.10 &    0.10 &     \nl 
 D33J013336.7+303128.7 &  4.078  &   21.45 &   21.53 &    0.26 &    0.18 &     \nl 
 D33J013357.9+303526.7 &  4.078  & \nodata &   21.34 & \nodata &    0.12 &     \nl 
 D33J013337.3+303425.1 &  4.099  &   21.84 & \nodata &    0.20 & \nodata &     \nl 
 D33J013357.3+303455.8 &  4.099  &   21.55 &   22.42 &    0.14 &    0.18 &     \nl 
 D33J013346.7+303149.7 &  4.121: &   21.35 &   22.07 &    0.13 &    0.24 &     \nl 
 D33J013355.9+303550.1 &  4.132  &   21.80 & \nodata &    0.34 & \nodata &     \nl 
 D33J013345.2+303506.3 &  4.143: &   21.59 &   22.64 &    0.19 &    0.40 &     \nl 
 D33J013352.0+303137.2 &  4.187  & \nodata &   23.11 & \nodata &    0.76 &     \nl 
 D33J013340.1+303642.3 &  4.233: &   22.62 &   23.21 &    0.28 &    0.44 &     \nl 
 D33J013336.4+303235.6 &  4.256: &   20.88 &   21.63 &    0.12 &    0.13 &     \nl 
 D33J013340.8+302920.2 &  4.256: &   21.76 &   22.47 &    0.15 &    0.31 &     \nl 
 D33J013342.1+303451.2 &  4.256  &   21.86 &   22.39 &    0.27 &    0.34 &     \nl 
 D33J013337.4+303221.9 &  4.279  &   21.36 &   22.20 &    0.14 &    0.18 &     \nl 
 D33J013348.8+303312.4 &  4.327  &   21.44 &   22.05 &    0.21 &    0.29 &     \nl 
 D33J013400.6+303556.2 &  4.327  &   21.79 &   22.47 &    0.33 &    0.57 &     \nl 
 D33J013405.7+303528.5 &  4.327  &   20.97 &   21.47 &    0.15 &    0.19 &     \nl 
 D33J013413.2+303653.8 &  4.327  &   21.45 &   22.07 &    0.16 &    0.25 &     \nl 
 D33J013337.1+303740.6 &  4.351  &   21.05 & \nodata &    0.14 & \nodata &     \nl 
 D33J013352.9+303202.6 &  4.351  &   21.79 &   22.62 &    0.26 &    0.49 &     \nl 
 D33J013354.3+303159.3 &  4.351  &   21.46 &   22.07 &    0.21 &    0.29 &     \nl 
 D33J013336.9+303709.3 &  4.360  &   20.88 &   21.80 &    0.21 &    0.37 & 1   \nl 
 D33J013336.0+303305.0 &  4.375: &   21.03 &   21.42 &    0.14 &    0.16 &     \nl 
 D33J013406.2+303842.6 &  4.375  &   21.12 &   21.29 &    0.18 &    0.17 &     \nl 
 D33J013414.8+303033.3 &  4.400  &   21.50 &   22.29 &    0.36 &    0.49 &     \nl 
 D33J013332.3+303046.7 &  4.425: &   21.94 & \nodata &    0.21 & \nodata &     \nl 
 D33J013354.2+303205.5 &  4.425  &   22.16 &   22.81 &    0.42 &    0.57 &     \nl 
 D33J013333.6+303002.9 &  4.450: &   21.93 &   22.64 &    0.28 &    0.46 &     \nl 
 D33J013352.9+303356.7 &  4.450: &   21.25 &   21.89 &    0.13 &    0.13 &     \nl 
 D33J013353.6+303001.3 &  4.450  &   21.50 &   22.27 &    0.29 &    0.36 &     \nl 
 D33J013415.5+302942.3 &  4.502: &   21.57 &   22.61 &    0.24 &    0.61 &     \nl 
 D33J013341.7+302928.2 &  4.528: &   21.94 &   22.57 &    0.22 &    0.31 &     \nl 
 D33J013350.4+303216.3 &  4.528  &   21.62 &   22.06 &    0.20 &    0.23 &     \nl 
 D33J013354.4+303523.6 &  4.528  &   21.45 &   22.04 &    0.14 &    0.22 &     \nl 
 D33J013332.6+303526.1 &  4.581  &   20.89 &   21.55 &    0.12 &    0.19 &     \nl 
 D33J013336.4+303645.7 &  4.608  & \nodata &   22.61 & \nodata &    0.50 &     \nl 
 D33J013338.8+303819.2 &  4.608  &   20.83 &   21.38 &    0.14 &    0.17 &     \nl 
 D33J013413.2+303315.1 &  4.608  &   21.48 & \nodata &    0.16 & \nodata &     \nl 
 D33J013411.3+303823.9 &  4.636  &   22.22 &   22.68 &    0.44 &    0.46 &     \nl 
 D33J013414.9+303612.8 &  4.636  &   21.48 &   21.96 &    0.17 &    0.22 &     \nl 
 D33J013342.6+303104.9 &  4.692: &   21.27 &   21.83 &    0.18 &    0.18 &     \nl 
 D33J013355.7+303351.5 &  4.692: &   21.44 &   22.25 &    0.28 &    0.46 &     \nl 
 D33J013411.1+303853.7 &  4.720: &   21.37 &   22.43 &    0.14 &    0.26 &     \nl 
 D33J013350.9+303117.2 &  4.749  &   21.54 &   22.36 &    0.29 &    0.39 &     \nl 
 D33J013352.0+303129.4 &  4.749  &   20.83 &   21.28 &    0.09 &    0.11 &     \nl 
 D33J013356.2+303227.7 &  4.772  &   21.49 &   22.17 &    0.26 &    0.36 &     \nl 
 D33J013337.5+303147.7 &  4.808: &   21.46 &   22.13 &    0.19 &    0.20 &     \nl 
 D33J013343.3+303416.3 &  4.808  &   21.74 &   22.84 &    0.16 &    0.44 &     \nl 
 D33J013345.5+303806.9 &  4.808: &   21.81 &   22.19 &    0.29 &    0.46 &     \nl 
 D33J013350.2+303212.1 &  4.808  &   21.64 &   22.38 &    0.26 &    0.37 &     \nl 
 D33J013339.1+303713.3 &  4.840  &   21.37 &   22.27 &    0.35 &    0.60 & 1   \nl 
 D33J013348.8+303449.8 &  4.898: &   21.52 &   21.89 &    0.16 &    0.14 &     \nl 
 D33J013349.9+303231.8 &  4.898  &   21.99 &   22.57 &    0.25 &    0.33 &     \nl 
 D33J013353.1+303251.9 &  4.898  &   21.95 &   22.38 &    0.32 &    0.32 &     \nl 
 D33J013358.6+303830.3 &  4.898  &   21.52 &   22.60 &    0.28 &    0.60 &     \nl 
 D33J013337.1+303060.0 &  4.929: &   20.58 &   21.14 &    0.07 &    0.11 &     \nl 
 D33J013342.9+302903.7 &  4.930  &   20.29 &   21.46 &    0.11 &    0.14 & 1   \nl 
 D33J013336.4+303437.8 &  4.980  &   21.43 &   22.09 &    0.34 &    0.34 & 1   \nl 
 D33J013335.1+303534.2 &  4.992  &   21.28 &   22.04 &    0.21 &    0.39 &     \nl 
 D33J013335.4+303138.3 &  4.992: &   21.98 &   22.63 &    0.19 &    0.21 &     \nl 
 D33J013331.6+303407.1 &  5.025  &   21.55 &   22.50 &    0.30 &    0.45 &     \nl 
 D33J013345.0+303001.3 &  5.025  &   21.17 &   21.70 &    0.14 &    0.15 &     \nl 
 D33J013351.0+303118.2 &  5.025  &   21.60 &   22.18 &    0.45 &    0.52 &     \nl 
 D33J013411.8+303743.1 &  5.025  &   21.84 & \nodata &    0.33 & \nodata &     \nl 
 D33J013333.2+303349.3 &  5.057  &   21.65 &   22.46 &    0.37 &    0.43 &     \nl 
 D33J013356.1+303223.1 &  5.090  &   21.46 &   22.32 &    0.30 &    0.51 &     \nl 
 D33J013403.9+303808.4 &  5.090  &   20.67 &   21.16 &    0.18 &    0.26 &     \nl 
 D33J013402.5+303601.0 &  5.124: &   21.71 &   22.09 &    0.23 &    0.22 &     \nl 
 D33J013417.5+303153.4 &  5.124  &   21.59 &   22.11 &    0.11 &    0.09 &     \nl 
 D33J013352.3+303217.6 &  5.158  &   21.10 &   21.52 &    0.19 &    0.23 &     \nl 
 D33J013355.5+303527.3 &  5.158: &   21.42 &   21.90 &    0.25 &    0.36 &     \nl 
 D33J013414.2+303530.6 &  5.158  &   22.50 & \nodata &    0.45 & \nodata &     \nl 
 D33J013358.1+302958.6 &  5.240  &   21.09 &   21.88 &    0.20 &    0.27 & 1   \nl 
 D33J013405.9+303453.8 &  5.280  &   21.04 &   21.60 &    0.33 &    0.49 & 1   \nl 
 D33J013342.6+303329.5 &  5.310  &   20.66 &   21.30 &    0.14 &    0.15 & 1   \nl 
 D33J013408.6+303754.8 &  5.320  &   21.13 &   21.90 &    0.31 &    0.51 & 1   \nl 
 D33J013340.8+303434.3 &  5.330  &   21.26 &   21.47 &    0.40 &    0.63 & 1   \nl 
 D33J013335.9+303118.6 &  5.334: &   20.99 & \nodata &    0.17 & \nodata &     \nl 
 D33J013403.8+303830.5 &  5.334  &   21.04 &   22.02 &    0.30 &    0.49 &     \nl 
 D33J013351.8+303450.0 &  5.371: &   21.48 &   22.26 &    0.29 &    0.45 &     \nl 
 D33J013354.3+303530.7 &  5.371  &   20.58 &   21.21 &    0.09 &    0.20 &     \nl 
 D33J013355.7+303711.4 &  5.371  &   21.56 &   22.62 &    0.22 &    0.77 &     \nl 
 D33J013342.5+302958.7 &  5.390  &   21.55 &   22.37 &    0.24 &    0.44 & 1   \nl 
 D33J013353.8+303212.0 &  5.408  &   21.14 &   21.41 &    0.21 &    0.20 &     \nl 
 D33J013333.4+303118.2 &  5.446: &   22.33 & \nodata &    0.32 & \nodata &     \nl 
 D33J013349.5+303501.4 &  5.446: &   21.27 &   22.06 &    0.19 &    0.31 &     \nl 
 D33J013359.5+303846.8 &  5.450  &   21.05 &   21.79 &    0.32 &    0.48 & 1   \nl 
 D33J013416.2+303752.2 &  5.485: & \nodata &   21.21 & \nodata &    0.13 &     \nl 
 D33J013351.3+303156.6 &  5.524  &   21.87 &   22.36 &    0.47 &    0.40 &     \nl 
 D33J013400.5+303546.1 &  5.524: &   22.06 &   23.05 &    0.25 &    0.50 &     \nl 
 D33J013403.0+303805.4 &  5.531  &   21.19 &   21.88 &    0.26 &    0.40 &     \nl 
 D33J013331.8+303727.2 &  5.540  &   21.93 &   22.51 &    0.42 &    0.55 & 1   \nl 
 D33J013357.0+303117.5 &  5.550  &   20.54 &   21.30 &    0.11 &    0.15 & 1   \nl 
 D33J013342.1+303712.5 &  5.563: &   21.89 &   23.14 &    0.19 &    0.24 &     \nl 
 D33J013351.2+303001.0 &  5.600  &   20.47 &   21.12 &    0.15 &    0.19 & 1   \nl 
 D33J013347.2+303622.0 &  5.603: &   20.89 &   21.39 &    0.12 &    0.15 &     \nl 
 D33J013406.2+303559.2 &  5.603  &   21.27 &   21.77 &    0.23 &    0.25 &     \nl 
 D33J013343.1+303754.3 &  5.644: & \nodata &   21.39 & \nodata &    0.09 &     \nl 
 D33J013354.3+303215.9 &  5.720  & \nodata &   21.25 & \nodata &    0.18 &     \nl 
 D33J013336.0+303306.2 &  5.769  &   20.70 &   21.45 &    0.11 &    0.21 &     \nl 
 D33J013411.7+303504.7 &  5.769  &   21.71 &   22.12 &    0.22 &    0.19 &     \nl 
 D33J013332.2+303001.9 &  5.790  &   21.20 &   21.87 &    0.34 &    0.40 & 1   \nl 
 D33J013333.8+303415.5 &  5.856: &   21.45 & \nodata &    0.17 & \nodata &     \nl 
 D33J013345.1+303838.5 &  5.856  &   20.85 &   21.07 &    0.30 &    0.25 & 1   \nl 
 D33J013350.9+303156.3 &  5.890  &   21.05 &   21.23 &    0.33 &    0.35 & 1   \nl 
 D33J013405.0+303557.5 &  5.890  &   21.10 &   21.88 &    0.29 &    0.52 & 1   \nl 
 D33J013407.3+303048.6 &  5.900  &   21.76 &   22.46 &    0.45 &    0.53 & 1   \nl 
 D33J013407.9+303831.6 &  5.900  &   20.60 &   21.26 &    0.22 &    0.33 & 1   \nl 
 D33J013335.8+303300.2 &  5.900: &   21.25 &   21.86 &    0.13 &    0.11 &     \nl 
 D33J013355.5+303152.3 &  5.900: &   21.56 &   22.43 &    0.15 &    0.21 &     \nl 
 D33J013356.5+303232.8 &  5.900: &   21.18 &   21.68 &    0.11 &    0.10 &     \nl 
 D33J013353.1+303217.5 &  5.945: &   21.87 &   22.81 &    0.41 &    0.64 &     \nl 
 D33J013359.8+303800.0 &  5.990  &   21.12 &   21.86 &    0.38 &    0.64 & 1   \nl 
 D33J013341.1+303453.8 &  5.991  &   21.18 &   21.70 &    0.30 &    0.36 &     \nl 
 D33J013350.6+303445.8 &  6.030  &   21.13 &   21.96 &    0.26 &    0.43 & 1   \nl 
 D33J013354.4+303222.7 &  6.085  &   21.13 &   21.52 &    0.26 &    0.37 &     \nl 
 D33J013402.9+303907.6 &  6.085  &   21.50 &   22.78 &    0.28 &    0.70 & 2   \nl 
 D33J013417.0+302923.1 &  6.110  &   21.23 &   21.94 &    0.51 &    0.54 & 1   \nl 
 D33J013406.2+303031.3 &  6.180  &   21.27 &   21.98 &    0.19 &    0.18 & 1   \nl 
 D33J013334.9+303735.3 &  6.181: &   21.74 &   22.51 &    0.30 &    0.27 &     \nl 
 D33J013352.3+303801.3 &  6.181  &   21.05 & \nodata &    0.33 & \nodata &     \nl 
 D33J013404.9+303630.9 &  6.181  &   20.35 &   21.13 &    0.10 &    0.18 &     \nl 
 D33J013402.9+303813.0 &  6.231  &   21.24 &   21.84 &    0.22 &    0.28 &     \nl 
 D33J013417.0+303415.5 &  6.250  &   21.14 &   21.88 &    0.30 &    0.43 & 1   \nl 
 D33J013400.1+303904.2 &  6.281  &   21.10 &   21.87 &    0.35 &    0.51 & 2   \nl 
 D33J013343.2+303148.6 &  6.332: &   21.54 &   21.98 &    0.35 &    0.36 &     \nl 
 D33J013400.5+303630.8 &  6.332  &   21.00 &   21.77 &    0.28 &    0.45 &     \nl 
 D33J013408.0+303845.7 &  6.332: &   20.95 &   21.28 &    0.14 &    0.15 &     \nl 
 D33J013335.6+303129.9 &  6.384  &   20.70 &   21.05 &    0.23 &    0.20 &     \nl 
 D33J013343.2+303243.2 &  6.436  &   20.66 &   20.93 &    0.20 &    0.19 &     \nl 
 D33J013416.6+303858.5 &  6.436: &   21.95 &   23.00 &    0.41 &    0.62 &     \nl 
 D33J013346.3+303626.7 &  6.490: &   20.58 &   21.18 &    0.07 &    0.09 &     \nl 
 D33J013351.8+303310.6 &  6.490: &   21.63 &   22.49 &    0.27 &    0.35 &     \nl 
 D33J013333.2+303326.4 &  6.545: &   21.00 &   22.09 &    0.12 &    0.16 &     \nl 
 D33J013400.4+303251.6 &  6.545: &   21.65 &   22.53 &    0.43 &    0.58 &     \nl 
 D33J013410.9+303845.1 &  6.545: &   21.58 &   22.66 &    0.41 &    0.67 & 2   \nl 
 D33J013338.5+303745.3 &  6.714: &   20.99 &   21.95 &    0.11 &    0.18 &     \nl 
 D33J013403.8+303731.9 &  6.714: &   20.62 &   21.34 &    0.10 &    0.11 &     \nl 
 D33J013404.5+303416.6 &  6.772  &   21.25 &   22.03 &    0.31 &    0.51 &     \nl 
 D33J013349.4+303009.4 &  6.780  &   20.95 &   21.60 &    0.30 &    0.36 & 1   \nl 
 D33J013407.7+303851.1 &  6.832  &   21.37 &   22.13 &    0.29 &    0.42 &     \nl 
 D33J013336.6+303416.7 &  6.954: &   21.19 &   22.07 &    0.21 &    0.50 &     \nl 
 D33J013402.8+303644.7 &  6.954  &   20.71 &   21.20 &    0.14 &    0.15 &     \nl 
 D33J013355.7+303903.6 &  7.016: &   20.99 &   21.39 &    0.12 &    0.14 & 2   \nl 
 D33J013332.5+303408.9 &  7.060  &   20.81 &   21.65 &    0.24 &    0.54 & 1   \nl 
 D33J013333.8+303427.8 &  7.060  &   20.73 &   21.55 &    0.18 &    0.30 & 1   \nl 
 D33J013337.9+303354.6 &  7.145: &   21.06 &   21.84 &    0.26 &    0.34 &     \nl 
 D33J013413.6+303027.7 &  7.300  &   21.22 &   22.01 &    0.23 &    0.17 & 1   \nl 
 D33J013335.9+303804.3 &  7.347  &   20.90 &   21.77 &    0.29 &    0.44 &     \nl 
 D33J013359.0+303756.3 &  7.347  &   21.28 &   22.22 &    0.55 &    0.94 & 1   \nl 
 D33J013356.1+303803.9 &  7.350  &   21.32 &   22.18 &    0.35 &    0.70 & 1   \nl 
 D33J013347.4+303848.5 &  7.417  &   20.53 & \nodata &    0.17 & \nodata & 2   \nl 
 D33J013351.3+303227.1 &  7.417  &   21.16 &   21.75 &    0.26 &    0.33 &     \nl 
 D33J013334.1+303311.0 &  7.489  &   21.67 &   22.53 &    0.29 &    0.30 &     \nl 
 D33J013336.5+303053.2 &  7.590  &   20.88 &   21.62 &    0.44 &    0.43 & 1   \nl 
 D33J013353.2+303506.1 &  7.629: &   21.20 &   22.01 &    0.24 &    0.38 &     \nl 
 D33J013347.9+302943.6 &  7.630  &   21.11 &   21.96 &    0.29 &    0.38 & 1   \nl 
 D33J013356.5+303547.6 &  7.635: &   21.49 &   22.30 &    0.16 &    0.32 &     \nl 
 D33J013410.8+303834.8 &  7.635  &   21.50 &   22.41 &    0.27 &    0.51 &     \nl 
 D33J013346.0+303747.0 &  7.711: &   20.11 &   20.83 &    0.04 &    0.07 &     \nl 
 D33J013358.6+303316.6 &  7.711: &   20.73 &   21.36 &    0.30 &    0.46 &     \nl 
 D33J013411.6+303255.0 &  7.711  &   21.12 & \nodata &    0.26 & \nodata &     \nl 
 D33J013417.2+303726.1 &  7.720  &   21.82 &   22.77 &    0.50 &    0.82 & 1   \nl 
 D33J013352.5+303219.3 &  7.747  &   20.88 &   22.03 &    0.19 &    0.45 &     \nl 
 D33J013350.7+303203.7 &  7.770  & \nodata &   20.62 & \nodata &    0.11 & 1   \nl 
 D33J013417.3+303211.5 &  7.960  &   20.94 &   21.38 &    0.40 &    0.31 & 1   \nl 
 D33J013332.4+303143.3 &  7.970  &   21.43 &   22.16 &    0.23 &    0.38 & 1   \nl 
 D33J013348.8+303415.8 &  7.970  &   20.56 &   21.20 &    0.27 &    0.41 & 1   \nl 
 D33J013405.4+303825.0 &  8.029: &   21.07 &   21.65 &    0.24 &    0.26 &     \nl 
 D33J013336.3+303243.7 &  8.180  &   21.24 &   22.03 &    0.23 &    0.29 & 1   \nl 
 D33J013348.8+303045.0 &  8.180  &   21.11 &   21.91 &    0.15 &    0.21 & 1   \nl 
 D33J013402.4+303831.8 &  8.330  &   20.84 &   21.25 &    0.22 &    0.23 & 1   \nl 
 D33J013339.8+303412.2 &  8.370  &   20.83 &   21.69 &    0.33 &    0.49 & 1   \nl 
 D33J013338.9+303504.2 &  8.465: &   20.83 &   21.44 &    0.28 &    0.34 &     \nl 
 D33J013406.6+303816.8 &  8.580  &   20.55 &   21.29 &    0.29 &    0.44 & 1   \nl 
 D33J013401.6+303858.2 &  8.653: &   20.54 &   21.37 &    0.39 &    0.60 & 2   \nl 
 D33J013343.5+303121.5 &  8.751: &   20.85 &   21.64 &    0.16 &    0.22 &     \nl 
 D33J013337.5+303305.1 &  8.980  &   21.27 &   21.84 &    0.25 &    0.39 & 1   \nl 
 D33J013339.0+303413.5 &  9.056  &   20.79 &   21.39 &    0.22 &    0.31 &     \nl 
 D33J013336.8+303434.4 &  9.060  &   20.41 &   21.13 &    0.25 &    0.42 & 1   \nl 
 D33J013413.9+303212.3 &  9.090  &   20.83 &   21.55 &    0.26 &    0.31 & 1   \nl 
 D33J013346.3+302908.9 &  9.120  &   20.53 &   21.02 &    0.14 &    0.10 & 1   \nl 
 D33J013409.3+302956.8 &  9.162: &   20.76 &   21.64 &    0.19 &    0.25 &     \nl 
 D33J013352.7+303416.2 &  9.220  &   21.13 &   22.17 &    0.24 &    0.49 & 1   \nl 
 D33J013345.9+303749.5 &  9.272  &   20.93 &   21.83 &    0.24 &    0.32 &     \nl 
 D33J013359.7+303720.9 &  9.383  &   20.64 &   21.30 &    0.13 &    0.15 &     \nl 
 D33J013414.2+303713.8 &  9.498  &   21.04 &   21.55 &    0.25 &    0.27 &     \nl 
 D33J013343.1+303648.9 &  9.590  &   20.67 &   21.48 &    0.23 &    0.32 & 1   \nl 
 D33J013402.1+303741.9 &  9.615  &   20.98 &   21.33 &    0.31 &    0.38 &     \nl 
 D33J013349.2+303218.1 &  9.810  &   21.07 &   21.94 &    0.20 &    0.27 & 1   \nl 
 D33J013333.2+303344.5 &  9.985: &   20.42 &   21.08 &    0.14 &    0.22 &     \nl 
 D33J013355.0+303537.0 & 10.120  &   20.62 &   21.38 &    0.33 &    0.43 & 1   \nl 
 D33J013342.1+303210.7 & 10.380  &   20.73 &   21.96 &    0.25 &    0.87 & 1   \nl 
 D33J013356.1+303903.0 & 10.430  &   20.56 &   21.52 &    0.32 &    0.54 & 1   \nl 
 D33J013409.5+303621.6 & 10.470  &   20.57 &   21.46 &    0.27 &    0.58 & 1   \nl 
 D33J013338.8+303422.6 & 10.600  &   20.47 &   20.93 &    0.25 &    0.26 & 1   \nl 
 D33J013341.9+302951.8 & 10.600  &   20.99 &   22.06 &    0.19 &    0.54 & 1   \nl 
 D33J013335.6+303649.2 & 10.700  &   20.94 &   21.73 &    0.30 &    0.42 & 1   \nl 
 D33J013415.4+303727.6 & 11.280  &   21.19 &   21.89 &    0.43 &    0.35 & 1   \nl 
 D33J013411.3+303535.2 & 11.450  &   20.90 &   21.82 &    0.39 &    0.56 & 1   \nl 
 D33J013353.4+303308.5 & 11.490  &   21.24 &   22.07 &    0.43 &    0.62 & 1   \nl 
 D33J013335.5+303330.2 & 11.520  &   20.51 &   21.34 &    0.41 &    0.52 & 1   \nl 
 D33J013413.4+303317.7 & 11.520  &   21.20 &   22.16 &    0.31 &    0.42 & 1   \nl 
 D33J013336.3+303730.7 & 11.790  &   21.06 &   21.98 &    0.31 &    0.31 & 1   \nl 
 D33J013337.7+303218.9 & 11.880  &   20.88 &   21.66 &    0.19 &    0.27 & 1   \nl 
 D33J013357.6+303805.4 & 12.340  &   20.48 &   21.10 &    0.42 &    0.69 & 1   \nl 
 D33J013349.8+303758.7 & 12.910  &   20.39 &   21.32 &    0.44 &    0.69 & 1   \nl 
 D33J013350.0+303014.9 & 12.970  &   20.61 &   21.43 &    0.31 &    0.37 & 1   \nl 
 D33J013406.1+303734.0 & 13.020  &   20.89 &   21.93 &    0.18 &    0.33 & 1   \nl 
 D33J013411.9+302947.6 & 13.310  &   20.58 &   21.60 &    0.43 &    0.47 & 1   \nl 
 D33J013412.0+303519.1 & 13.370  &   20.83 &   21.73 &    0.55 &    0.78 & 1   \nl 
 D33J013351.2+303758.2 & 13.560  &   20.14 &   20.98 &    0.33 &    0.48 & 1   \nl 
 D33J013402.5+303628.0 & 13.660  &   20.50 &   21.20 &    0.33 &    0.49 & 1   \nl 
 D33J013331.6+303704.5 & 13.760  &   20.87 &   21.97 &    0.29 &    0.30 & 1   \nl 
 D33J013338.8+303751.1 & 13.780  &   21.00 &   22.13 &    0.38 &    0.45 & 1   \nl 
 D33J013336.5+302933.5 & 13.940  &   21.27 &   22.51 &    0.20 &    0.51 & 1   \nl 
\enddata 
\tablecomments{(1) Variables identified by Macri et al. (2001a)\\
(2) Variables identified by Mochejska et al. (2001a)} 
\label{tab:ceph}
\end{planotable} 
\end{small}

\begin{small}
\tablenum{5}
\begin{planotable}{lrl}
\tablewidth{0pc}
\tablecaption{\sc DIRECT Flux Cepheids in M33B}
\tablehead{
\colhead{} & \colhead{$P$} & \colhead{} \\
\colhead{Name} & \colhead{$(days)$} & \colhead{Comments}}
\startdata
 D33J013401.3+303907.9 &  3.218  &     \nl 
 D33J013355.0+303214.3 &  4.099  &     \nl 
 D33J013402.3+303855.0 &  4.210  & 2   \nl 
 D33J013408.5+303323.5 &  4.375: &     \nl 
 D33J013357.1+303212.8 &  4.635  &     \nl 
 D33J013338.6+303757.4 &  4.692: &     \nl 
 D33J013416.5+303223.1 &  4.749  &     \nl 
 D33J013351.0+303145.3 &  4.898  &     \nl 
 D33J013408.6+303831.5 &  4.898  &     \nl 
 D33J013415.4+303715.9 &  5.298: &     \nl 
 D33J013402.5+303845.0 &  6.132  &     \nl 
 D33J013353.6+303151.4 &  6.231: &     \nl 
 D33J013416.7+303559.4 &  6.490: &     \nl 
 D33J013336.4+303228.4 &  7.211: &     \nl 
 D33J013353.4+303757.5 &  7.347: &     \nl 
 D33J013332.6+303115.1 &  7.867: &     \nl 
 D33J013357.3+303840.1 &  9.735  & 2   \nl 
 D33J013412.5+303839.8 & 11.240  & 1   \nl 
 D33J013339.5+303416.9 & 13.907  &     \nl 
 D33J013346.1+303809.3 & 14.422  &     \nl 
\enddata
\tablecomments{(1) Variables identified by Macri et al. (2001a)\\
(2) Variables identified by Mochejska et al. (2001a)}
\label{tab:ceph_flux}
\end{planotable}
\end{small}

\begin{small}
\tablenum{6}
\begin{planotable}{ccccc}
\tablewidth{0pc}
\tablecaption{\sc Light Curves of Cepheids in M33B}
\tablehead{ \colhead{Name} & \colhead{Filter} &
\colhead{HJD$-$2451000} & \colhead{flux} & \colhead{$\sigma_{flux}$} }
\startdata
D33J013402.3+303855.0 & b & 452.7982 &    -78.865 & 116.725\\
                      & b & 452.9166 &     71.067 & 157.887\\
                      & b & 454.6968 &    395.685 & 122.487\\
                      & b & 454.7299 &    335.538 &  91.812\\
\enddata
\tablecomments{Table 6 is available in its entirety in the electronic
version of the Astronomical Journal. A portion is shown here for
guidance regarding its form and content.}
\label{tab:lc_ceph}
\end{planotable}
\end{small}

\begin{small}
\tablenum{7}
\begin{planotable}{lrrrrrl}
\tablewidth{0pc}
\tablecaption{\sc DIRECT Other Periodic Variables in M33B}
\tablehead{
\colhead{} & \colhead{$P$} & \colhead{} & \colhead{} & \colhead{} & 
\colhead{} & \colhead{} \\
\colhead{Name} & \colhead{$(days)$} & \colhead{$V^a$} & \colhead{$B^a$} & 
\colhead{$A_V$} & \colhead{$A_B$} & \colhead{Comments}}
\startdata
 D33J013340.2+303722.2 &  1.98  &   19.41 &   19.32 &    0.12 &    0.07 &      EB\nl 
 D33J013338.5+303113.3 &  2.01  &   20.82 &   20.61 &    0.58 &    0.42 &      EB\nl 
 D33J013359.8+303354.9 &  2.11  &   17.91 &   17.85 &    0.10 &    0.12 &       \nl 
 D33J013407.7+303454.7 &  2.18  &   19.14 &   19.06 &    0.03 &    0.04 &       \nl 
 D33J013340.7+303054.5 &  2.32  &   20.55 &   20.54 &    0.06 &    0.03 &       \nl 
 D33J013354.4+303357.4 &  2.78: &   20.76 &   20.93 &    0.08 &    0.08 &       \nl 
 D33J013410.6+303516.8 &  3.18  & \nodata &   20.50 & \nodata &    0.06 &       \nl 
 D33J013332.5+303335.3 &  3.19: &   19.44 &   19.26 &    0.10 &    0.05 &      EB\nl 
 D33J013334.8+303211.4 &  3.40  & \nodata & \nodata & \nodata & \nodata &      1\nl 
 D33J013356.9+303752.4 &  3.66: &   20.29 &   21.15 &    0.06 &    0.13 &       \nl 
 D33J013339.4+303124.6 &  3.69  &   16.98 & \nodata &    0.01 & \nodata &       \nl 
 D33J013334.2+303058.0 &  3.71  &   21.23 &   21.65 &    0.21 &    0.29 &      1\nl 
 D33J013416.9+303454.6 &  4.57  &   21.22 &   20.92 &    0.13 &    0.09 &      1\nl 
 D33J013359.1+303523.9 &  4.66  & \nodata &   20.75 & \nodata &    0.10 &       \nl 
 D33J013351.8+303159.7 &  4.73: &   20.65 &   20.73 &    0.09 &    0.08 &       \nl 
 D33J013354.8+303248.9 &  4.87  &   18.30 &   18.24 &    0.04 &    0.04 &      1\nl 
 D33J013339.7+302942.9 &  4.93  &   21.04 &   21.38 &    0.13 &    0.15 &      1\nl 
 D33J013341.6+303220.3 &  5.30  &   16.32 &   17.11 &    0.03 &    0.02 &      1\nl 
 D33J013353.7+303519.6 &  5.81: &   19.62 &   19.50 &    0.17 &    0.13 &      EB\nl 
 D33J013341.8+303452.1 &  5.97  &   21.22 &   21.74 &    0.32 &    0.39 &      1\nl 
 D33J013344.6+303145.5 &  6.08  &   19.42 &   19.40 &    0.03 &    0.04 &       \nl 
 D33J013412.8+303840.1 &  6.18  & \nodata &   19.87 & \nodata &    0.05 &       \nl 
 D33J013342.6+303603.3 &  6.23: &   18.63 &   18.52 &    0.06 &    0.04 &      EB\nl 
 D33J013356.5+303316.2 &  6.53  &   17.05 &   17.04 &    0.01 &    0.01 &       \nl 
 D33J013413.7+303551.8 &  7.39  &   20.74 &   20.73 &    0.43 &    0.41 &      EB\nl 
 D33J013352.0+303542.6 &  8.85: & \nodata &   19.36 & \nodata &    0.03 &       \nl 
 D33J013333.5+303320.5 &  9.01  & \nodata & \nodata & \nodata & \nodata &      1\nl 
 D33J013347.8+303813.7 &  9.38  & \nodata & \nodata & \nodata & \nodata &      1\nl 
 D33J013358.6+303241.8 &  9.62  &   19.79 &   20.06 &    0.14 &    0.15 &       \nl 
 D33J013401.4+303727.1 & 12.56: & \nodata &   20.14 & \nodata &    0.11 &       \nl 
\enddata
\tablecomments{
$^a$ The $V$ and $B$ columns list the maximum magnitudes $V_{max}$
and $B_{max}$ for the eclipsing variables and flux-weighted average
magnitudes $\langle V\rangle$ and $\langle B\rangle$ for the other
variables.\\
(1) Variables identified by Macri et al. (2001a)
}
\label{tab:per}
\end{planotable}
\end{small}

\begin{small}
\tablenum{8}
\begin{planotable}{ccccc}
\tablewidth{0pc}
\tablecaption{\sc Light Curves of Other Periodic Variables in M33B}
\tablehead{ \colhead{Name} & \colhead{Filter} &
\colhead{HJD$-$2451000} & \colhead{mag} & \colhead{$\sigma_{mag}$} }
\startdata
D33J013340.2+303722.2 & B & 452.7982  &    19.354 &   0.004\\
                      & B & 452.9166  &    19.337 &   0.006\\
                      & B & 454.6968  &    19.394 &   0.005\\
                      & B & 454.7299  &    19.357 &   0.004\\
\enddata
\tablecomments{Table 8 is available in its entirety in the electronic
version of the Astronomical Journal. A portion is shown here for
guidance regarding its form and content.}
\label{tab:lc_per}
\end{planotable}
\end{small}

\begin{small}
\tablenum{9}
\begin{planotable}{lrrrrl}
\tablewidth{0pc}
%\tablewidth{310pt}
\tablecaption{\sc DIRECT Miscellaneous Variables in M33B}
\tablehead{
\colhead{Name} & \colhead{$\bar{V}$} &
\colhead{$\bar{B}$} & \colhead{$A_V$} & \colhead{$A_B$} & \colhead{Comments}}
\startdata
 D33J013345.1+303619.8 &   16.56 &   17.11 &    0.06 &    0.05 &      1\nl 
 D33J013401.8+303858.3 &   16.65 &   18.06 &    0.03 &    0.04 &      1\nl 
 D33J013358.1+303320.7 &   16.67 &   16.78 &    0.07 &    0.05 &      LP\nl 
 D33J013352.5+303816.0 &   16.77 &   17.63 &    0.04 &    0.09 &      1\nl 
 D33J013338.1+303110.3 &   16.82 &   16.82 &    0.06 &    0.02 &      LP\nl 
 D33J013339.3+303118.6 &   16.88 &   18.62 &    0.01 &    0.05 &      1\nl 
 D33J013357.9+303302.4 &   16.89 & \nodata &    0.03 & \nodata &       \nl 
 D33J013410.9+303840.9 &   16.93 &   17.45 &    0.07 &    0.08 &       \nl 
 D33J013355.6+303500.7 &   17.00 &   17.56 &    0.11 &    0.12 &      LP\nl 
 D33J013351.6+303454.7 &   17.01 &   17.11 &    0.05 &    0.04 &      LP\nl 
 D33J013335.2+303559.9 &   17.08 &   17.20 &    0.03 &    0.05 &      1\nl 
 D33J013416.4+303120.9 &   17.14 &   17.23 &    0.15 &    0.10 &      LP\nl 
 D33J013344.2+303147.9 &   17.22 &   17.21 &    0.05 &    0.03 &      LP\nl 
 D33J013344.1+303205.6 &   17.23 &   18.20 &    0.01 &    0.04 &      1\nl 
 D33J013417.8+303327.1 &   17.25 &   17.23 &    0.06 &    0.05 &      LP\nl 
 D33J013416.9+303856.8 &   17.26 & \nodata &    0.06 & \nodata &       \nl 
 D33J013359.0+303353.7 &   17.27 &   17.33 &    0.06 &    0.04 &      LP\nl 
 D33J013333.4+303407.3 &   17.28 &   17.34 &    0.03 &    0.03 &      LP\nl 
 D33J013357.5+303306.4 &   17.30 & \nodata &    0.04 & \nodata &      LP\nl 
 D33J013346.3+303257.3 &   17.34 &   17.57 &    0.04 &    0.03 &       \nl 
 D33J013345.0+303616.8 &   17.42 &   17.41 &    0.05 &    0.06 &      LP\nl 
 D33J013333.2+303505.7 &   17.52 &   17.60 &    0.03 &    0.03 &      1\nl 
 D33J013400.9+303414.9 &   17.55 &   19.26 &    0.06 &    0.07 &      1\nl 
 D33J013354.6+303308.1 &   17.56 & \nodata &    0.08 & \nodata &      LP\nl 
 D33J013409.2+303423.2 &   17.64 &   18.45 &    0.05 &    0.08 &      1\nl 
 D33J013350.6+303230.3 &   17.70 & \nodata &    0.11 & \nodata &      1\nl 
 D33J013335.7+303842.8 &   17.79 & \nodata &    0.10 & \nodata &      LP\nl 
 D33J013352.1+303636.5 &   17.79 &   18.03 &    0.04 &    0.04 &      LP\nl 
 D33J013359.8+303427.1 &   17.79 &   17.73 &    0.05 &    0.03 &       \nl 
 D33J013332.4+303543.3 &   17.80 &   17.73 &    0.05 &    0.07 &      LP\nl 
 D33J013347.3+303306.5 &   17.80 &   17.84 &    0.05 &    0.06 &      LP\nl 
 D33J013351.1+303811.1 &   17.80 & \nodata &    0.10 & \nodata &      LP\nl 
 D33J013344.4+303843.9 &   17.83 &   17.99 &    0.04 &    0.04 &      LP\nl 
 D33J013401.0+303634.7 &   17.84 &   17.99 &    0.04 &    0.02 &      LP\nl 
 D33J013409.2+303853.1 &   17.85 &   18.12 &    0.11 &    0.08 &      LP\nl 
 D33J013359.2+303535.2 &   17.89 &   17.91 &    0.04 &    0.06 &      LP\nl 
 D33J013359.6+303333.0 &   17.91 &   18.11 &    0.04 &    0.03 &      LP\nl 
 D33J013416.1+303641.8 &   17.91 &   17.93 &    0.06 &    0.06 &      LP\nl 
 D33J013338.2+303818.9 &   17.93 &   17.93 &    0.07 &    0.04 &       \nl 
 D33J013351.4+303848.8 &   17.93 &   18.03 &    0.09 &    0.08 &      LP\nl 
 D33J013344.8+303217.5 &   17.94 &   18.04 &    0.03 &    0.03 &      LP\nl 
 D33J013409.6+303638.2 &   17.94 &   17.95 &    0.07 &    0.05 &      LP\nl 
 D33J013352.8+303819.4 &   17.97 &   17.98 &    0.09 &    0.08 &      LP\nl 
 D33J013358.5+303419.8 &   17.99 &   19.69 &    0.05 &    0.05 &      1\nl 
 D33J013415.2+303659.1 &   18.01 &   18.22 &    0.06 &    0.04 &      LP\nl 
 D33J013350.4+303855.8 &   18.07 &   18.18 &    0.04 &    0.05 &      LP\nl 
 D33J013350.4+303817.1 &   18.10 &   18.07 &    0.08 &    0.09 &      LP\nl 
 D33J013341.3+303212.7 &   18.11 & \nodata &    0.15 & \nodata &      1\nl 
 D33J013406.7+303631.4 &   18.18 &   18.23 &    0.04 &    0.03 &      LP\nl 
 D33J013414.9+303436.4 &   18.28 &   19.09 &    0.10 &    0.15 &       \nl 
 D33J013343.3+303318.6 &   18.30 & \nodata &    0.09 & \nodata &      1\nl 
 D33J013359.5+303021.8 &   18.30 &   18.34 &    0.11 &    0.09 &      LP\nl 
 D33J013349.9+302928.9 &   18.33 &   18.96 &    0.10 &    0.04 &      1\nl 
 D33J013414.6+303326.6 &   18.36 &   19.02 &    0.06 &    0.06 &       \nl 
 D33J013344.8+303210.8 &   18.41 &   18.58 &    0.03 &    0.02 &      LP\nl 
 D33J013352.3+303746.2 &   18.44 &   18.53 &    0.06 &    0.06 &      LP\nl 
 D33J013350.3+303226.0 &   18.46 &   18.51 &    0.06 &    0.04 &      LP\nl 
 D33J013345.2+303138.2 &   18.50 &   20.36 &    0.12 &    0.15 &      1\nl 
 D33J013340.1+303549.8 &   18.52 &   18.57 &    0.07 &    0.08 &      LP\nl 
 D33J013336.7+303531.8 &   18.53 &   20.38 &    0.04 &    0.08 &      1\nl 
 D33J013416.3+303353.4 &   18.58 &   20.46 &    0.26 &    0.20 &      1\nl 
 D33J013340.0+303507.5 &   18.60 &   19.41 &    0.54 &    0.69 &       \nl 
 D33J013357.2+303429.2 &   18.62 &   18.56 &    0.05 &    0.04 &      LP\nl 
 D33J013356.8+303529.7 &   18.66 &   18.75 &    0.03 &    0.04 &      LP\nl 
 D33J013401.6+303551.9 &   18.66 &   18.60 &    0.05 &    0.03 &      LP\nl 
 D33J013355.6+303334.2 &   18.67 &   18.84 &    0.03 &    0.05 &      LP\nl 
 D33J013349.8+303224.6 &   18.73 &   20.44 &    0.10 &    0.08 &      1\nl 
 D33J013339.9+303810.4 &   18.76 &   18.64 &    0.07 &    0.05 &      LP\nl 
 D33J013401.1+303010.9 &   18.77 &   18.73 &    0.07 &    0.04 &      LP\nl 
 D33J013344.3+303635.7 &   18.81 &   19.12 &    0.06 &    0.03 &      1\nl 
 D33J013357.8+303338.9 &   18.83 &   20.44 &    0.29 &    0.23 &       \nl 
 D33J013344.6+303552.7 &   18.84 &   18.92 &    0.06 &    0.06 &      LP\nl 
 D33J013341.2+303525.6 &   18.87 &   18.91 &    0.06 &    0.04 &      LP\nl 
 D33J013344.1+303600.9 &   18.89 &   18.81 &    0.10 &    0.07 &      LP\nl 
 D33J013400.5+303536.4 &   18.92 &   18.85 &    0.04 &    0.04 &      LP\nl 
 D33J013411.5+303312.6 &   19.03 &   20.40 &    0.12 &    0.08 &      1\nl 
 D33J013333.4+303350.7 &   19.05 &   20.98 &    0.09 &    0.11 &      1\nl 
 D33J013353.0+303842.8 &   19.06 &   19.06 &    0.06 &    0.04 &      1\nl 
 D33J013415.3+303633.5 &   19.06 &   18.93 &    0.10 &    0.07 &      LP\nl 
 D33J013343.9+303800.7 &   19.10 &   19.10 &    0.06 &    0.04 &      LP\nl 
 D33J013402.3+303828.3 &   19.21 &   21.00 &    0.08 &    0.08 &      1\nl 
 D33J013353.5+303519.9 &   19.24 & \nodata &    0.05 & \nodata &      1\nl 
 D33J013339.4+303512.4 &   19.25 &   20.93 &    0.04 &    0.06 &      1\nl 
 D33J013357.0+303355.3 &   19.26 &   21.22 &    0.06 &    0.12 &      1\nl 
 D33J013414.5+303557.7 &   19.28 &   21.16 &    0.05 &    0.08 &      1\nl 
 D33J013331.5+303410.2 &   19.32 &   21.04 &    0.04 &    0.14 &      1\nl 
 D33J013333.5+303149.2 &   19.32 &   20.95 &    0.10 &    0.10 &      1\nl 
 D33J013335.5+303403.2 &   19.34 &   20.50 &    0.07 &    0.07 &      1\nl 
 D33J013406.2+303808.1 &   19.35 &   19.64 &    0.09 &    0.09 &       \nl 
 D33J013335.2+303040.6 &   19.40 &   20.98 &    0.05 &    0.06 &      1\nl 
 D33J013338.8+303532.5 &   19.41 &   21.35 &    0.10 &    0.14 &      1\nl 
 D33J013342.4+303631.0 &   19.42 &   21.31 &    0.13 &    0.16 &      1\nl 
 D33J013412.2+303320.6 &   19.43 &   21.23 &    0.24 &    0.21 &      1\nl 
 D33J013335.5+303015.3 &   19.44 &   21.14 &    0.20 &    0.15 &       \nl 
 D33J013416.3+303158.6 &   19.44 &   21.14 &    0.05 &    0.08 &      1\nl 
 D33J013351.2+303510.3 &   19.47 &   21.21 &    0.12 &    0.17 &       \nl 
 D33J013356.2+303258.7 &   19.48 &   21.35 &    0.15 &    0.14 &      1\nl 
 D33J013333.9+303402.6 &   19.50 &   21.11 &    0.12 &    0.13 &      1\nl 
 D33J013350.5+303222.1 &   19.51 &   20.61 &    0.10 &    0.07 &       \nl 
 D33J013405.4+303632.4 &   19.52 &   19.77 &    0.11 &    0.12 &      LP\nl 
 D33J013403.6+303143.0 &   19.53 &   21.47 &    0.04 &    0.07 &      1\nl 
 D33J013406.7+303430.0 &   19.53 &   21.33 &    0.16 &    0.20 &       \nl 
 D33J013348.6+303247.8 &   19.56 &   19.79 &    0.10 &    0.10 &      LP\nl 
 D33J013414.5+303511.5 &   19.56 &   21.11 &    0.05 &    0.09 &      1\nl 
 D33J013339.0+303505.6 &   19.57 &   20.79 &    0.04 &    0.08 &      1\nl 
 D33J013353.6+303210.6 &   19.58 &   21.31 &    0.05 &    0.09 &      1\nl 
 D33J013352.1+303902.3 &   19.59 &   21.91 &    0.21 &    0.38 &       \nl 
 D33J013412.9+303309.9 &   19.59 &   21.49 &    0.06 &    0.08 &      1\nl 
 D33J013340.8+303236.1 &   19.62 &   21.40 &    0.11 &    0.11 &       \nl 
 D33J013332.0+303338.0 &   19.64 &   21.11 &    0.11 &    0.09 &       \nl 
 D33J013331.1+303502.4 &   19.68 &   21.19 &    0.08 &    0.13 &      1\nl 
 D33J013359.1+303212.2 &   19.69 &   21.70 &    0.08 &    0.16 &      1\nl 
 D33J013358.5+303812.8 &   19.72 &   21.46 &    0.17 &    0.14 &      1\nl 
 D33J013401.2+303557.3 &   19.72 &   21.29 &    0.09 &    0.12 &      1\nl 
 D33J013342.3+303608.0 &   19.73 &   21.58 &    0.17 &    0.15 &      1\nl 
 D33J013348.8+303709.2 &   19.73 &   21.37 &    0.10 &    0.15 &      1\nl 
 D33J013357.0+303516.9 &   19.73 &   21.79 &    0.06 &    0.16 &      1\nl 
 D33J013352.4+303840.2 &   19.75 &   20.01 &    0.11 &    0.13 & 2    LP\nl 
 D33J013343.5+302938.3 &   19.77 &   21.56 &    0.06 &    0.09 &      1\nl 
 D33J013350.6+303617.1 &   19.79 &   21.49 &    0.05 &    0.07 &      1\nl 
 D33J013351.4+303842.3 &   19.80 &   19.70 &    0.08 &    0.04 &      1\nl 
 D33J013400.5+302951.5 &   19.82 &   21.32 &    0.33 &    0.20 &      1\nl 
 D33J013335.9+303344.0 &   19.83 &   21.93 &    0.18 &    0.16 &      1\nl 
 D33J013344.4+303227.7 &   19.84 &   21.69 &    0.12 &    0.13 &      1\nl 
 D33J013357.4+303409.1 &   19.84 &   19.70 &    0.16 &    0.12 &      LP\nl 
 D33J013345.1+303606.3 &   19.85 &   21.16 &    0.14 &    0.19 &      1\nl 
 D33J013405.5+303443.2 &   19.86 &   21.57 &    0.09 &    0.13 &      1\nl 
 D33J013407.7+303742.4 &   19.86 &   20.06 &    0.27 &    0.28 &      LP\nl 
 D33J013342.6+303534.4 &   19.89 &   21.38 &    0.07 &    0.11 &      1\nl 
 D33J013352.4+303736.4 &   19.90 &   21.48 &    0.25 &    0.20 &      1\nl 
 D33J013359.4+303734.0 &   19.91 &   21.45 &    0.05 &    0.12 &      1\nl 
 D33J013401.6+303129.0 &   19.95 &   21.89 &    0.13 &    0.14 &      1\nl 
 D33J013401.5+303859.2 &   19.96 &   19.98 &    0.11 &    0.05 &      1\nl 
 D33J013359.5+303200.3 &   19.98 & \nodata &    0.11 & \nodata &       \nl 
 D33J013343.7+303134.1 &   19.99 &   20.22 &    0.14 &    0.13 &      LP\nl 
 D33J013415.5+303107.3 &   19.99 &   21.11 &    0.20 &    0.26 &      LP\nl 
 D33J013344.0+303609.5 &   20.01 & \nodata &    0.08 & \nodata &      LP\nl 
 D33J013416.3+303801.6 &   20.04 &   21.73 &    0.20 &    0.15 &      1\nl 
 D33J013345.7+303609.5 &   20.05 & \nodata &    0.22 & \nodata &       \nl 
 D33J013355.1+303109.9 &   20.07 &   20.62 &    0.14 &    0.18 &       \nl 
 D33J013334.4+303426.4 &   20.08 &   20.60 &    0.16 &    0.25 &      LP\nl 
 D33J013354.3+303320.5 &   20.08 &   21.29 &    0.07 &    0.09 &      1\nl 
 D33J013337.4+303752.5 &   20.09 &   20.72 &    0.18 &    0.18 &      LP\nl 
 D33J013352.2+303646.6 &   20.09 &   21.86 &    0.20 &    0.24 &      1\nl 
 D33J013339.0+303828.9 &   20.11 &   21.09 &    0.12 &    0.09 &      1\nl 
 D33J013416.4+303545.8 &   20.13 &   20.22 &    0.20 &    0.14 &      LP\nl 
 D33J013339.3+303049.4 &   20.14 &   20.41 &    0.04 &    0.04 &      1\nl 
 D33J013348.9+303826.6 &   20.22 & \nodata &    0.45 & \nodata &      LP\nl 
 D33J013350.5+303225.3 &   20.24 &   22.05 &    0.17 &    0.19 &      1\nl 
 D33J013338.4+303638.9 &   20.26 &   21.24 &    0.36 &    0.71 &      LP\nl 
 D33J013401.0+303432.2 &   20.31 &   22.05 &    0.09 &    0.16 &      1\nl 
 D33J013332.8+303247.0 &   20.32 &   20.14 &    0.27 &    0.16 &      LP\nl 
 D33J013354.6+303444.8 &   20.34 &   20.84 &    0.15 &    0.17 &       \nl 
 D33J013351.4+303640.0 &   20.36 &   22.38 &    0.22 &    0.27 &      1\nl 
 D33J013358.4+303429.4 &   20.42 &   21.16 &    0.58 &    1.16 &      LP\nl 
 D33J013338.0+303235.6 &   20.51 &   22.06 &    0.20 &    0.16 &      1\nl 
 D33J013406.1+303507.5 &   20.52 &   20.82 &    0.14 &    0.16 &      LP\nl 
 D33J013417.8+303355.9 &   20.52 &   21.83 &    0.22 &    0.15 &      1\nl 
 D33J013405.5+303418.9 &   20.53 &   20.91 &    0.10 &    0.11 &      1\nl 
 D33J013349.3+303159.4 &   20.54 &   20.59 &    0.19 &    0.12 &      LP\nl 
 D33J013409.7+303255.4 &   20.55 &   20.55 &    0.20 &    0.16 &       \nl 
 D33J013333.4+303146.7 &   20.58 &   22.39 &    0.08 &    0.16 &      1\nl 
 D33J013400.1+303755.8 &   20.59 &   20.61 &    0.22 &    0.14 &       \nl 
 D33J013415.1+303655.8 &   20.60 &   20.82 &    1.11 &    1.04 &      LP\nl 
 D33J013342.2+303640.7 &   20.62 &   21.17 &    0.24 &    0.12 &       \nl 
 D33J013416.5+303728.6 &   20.63 &   21.47 &    0.60 &    0.79 &      LP\nl 
 D33J013345.5+303521.3 &   20.67 &   22.03 &    0.19 &    0.18 &       \nl 
 D33J013340.4+303131.1 &   20.69 & \nodata &    0.11 & \nodata &      1\nl 
 D33J013357.7+303235.1 &   20.70 &   22.30 &    0.48 &    0.37 &      1\nl 
 D33J013405.6+303905.4 &   20.70 &   21.73 &    0.44 &    0.33 &       \nl 
 D33J013414.1+303609.4 &   20.74 &   21.32 &    0.22 &    0.31 &       \nl 
 D33J013345.2+303444.1 &   20.77 &   22.11 &    0.28 &    0.25 &       \nl 
 D33J013411.4+303125.8 &   20.78 &   20.68 &    0.21 &    0.13 &      LP\nl 
 D33J013401.2+303423.1 &   20.81 &   20.64 &    0.21 &    0.17 &       \nl 
 D33J013339.1+302944.1 &   20.82 &   22.57 &    0.15 &    0.45 &      1\nl 
 D33J013340.1+303136.5 &   20.86 &   21.49 &    0.74 &    0.61 &      LP\nl 
 D33J013331.3+303354.9 &   20.90 &   21.75 &    0.38 &    0.44 &      LP\nl 
 D33J013350.7+303844.6 &   20.91 &   20.87 &    0.35 &    0.12 &       \nl 
 D33J013356.1+302944.1 &   20.93 &   21.00 &    0.10 &    0.07 &      1\nl 
 D33J013359.7+303111.0 &   20.96 &   22.08 &    0.22 &    0.26 &       \nl 
 D33J013414.6+303723.0 &   20.96 &   22.40 &    0.25 &    0.29 &       \nl 
 D33J013405.9+303819.4 &   20.97 &   21.83 &    0.56 &    0.79 &      LP\nl 
 D33J013336.5+303650.2 &   21.01 &   22.36 &    0.56 &    1.24 &      LP\nl 
 D33J013351.6+303653.4 &   21.02 &   22.26 &    0.46 &    0.48 &       \nl 
 D33J013351.7+303815.8 &   21.02 &   21.99 &    0.23 &    0.25 &       \nl 
 D33J013407.5+303648.3 &   21.04 &   21.06 &    0.27 &    0.30 &       \nl 
 D33J013340.0+303201.1 &   21.05 & \nodata &    0.17 & \nodata &      1\nl 
 D33J013353.8+303815.2 &   21.05 &   22.03 &    0.72 &    0.47 &       \nl 
 D33J013404.3+303749.1 &   21.12 &   22.56 &    0.36 &    0.65 &       \nl 
 D33J013336.1+303458.6 &   21.17 &   23.10 &    0.52 &    0.83 &       \nl 
 D33J013345.6+303300.1 &   21.20 & \nodata &    0.14 & \nodata &      1\nl 
 D33J013411.2+303748.4 &   21.22 &   22.59 &    0.35 &    0.45 &       \nl 
 D33J013357.6+303249.5 &   21.23 &   21.24 &    0.31 &    0.22 &       \nl 
 D33J013358.6+303245.7 &   21.24 &   22.60 &    0.27 &    0.20 &       \nl 
 D33J013338.0+303540.0 &   21.26 &   22.96 &    0.58 &    0.68 &       \nl 
 D33J013341.7+303405.7 &   21.26 & \nodata &    0.34 & \nodata &       \nl 
 D33J013403.8+303753.0 &   21.27 &   22.31 &    0.29 &    0.23 &      1\nl 
 D33J013341.1+303742.9 &   21.29 &   23.14 &    0.29 &    0.65 &       \nl 
 D33J013356.6+303818.2 &   21.30 &   22.87 &    0.78 &    0.76 &       \nl 
 D33J013410.8+303146.5 &   21.31 &   22.18 &    0.30 &    0.28 &      LP\nl 
 D33J013338.5+302909.7 &   21.32 &   22.67 &    0.35 &    0.64 &       \nl 
 D33J013345.0+303700.8 &   21.34 & \nodata &    0.88 & \nodata &       \nl 
 D33J013347.5+303751.0 &   21.34 & \nodata &    0.24 & \nodata &       \nl 
 D33J013403.7+303045.5 &   21.35 &   21.49 &    0.32 &    0.33 &       \nl 
 D33J013335.8+303123.1 &   21.40 &   22.11 &    0.37 &    0.14 &       \nl 
 D33J013343.1+303747.7 &   21.45 &   22.52 &    0.33 &    0.25 &       \nl 
 D33J013357.5+303349.6 &   21.45 &   22.00 &    0.40 &    0.19 &       \nl 
 D33J013333.1+303314.1 &   21.47 &   22.79 &    0.62 &    0.37 &       \nl 
 D33J013338.2+303629.4 &   21.47 &   22.49 &    0.61 &    0.44 &       \nl 
 D33J013344.5+303813.8 &   21.47 & \nodata &    0.66 & \nodata &       \nl 
 D33J013401.5+303438.9 &   21.47 &   22.43 &    0.54 &    0.34 &       \nl 
 D33J013409.6+303559.5 &   21.47 &   22.93 &    0.45 &    0.54 &      LP\nl 
 D33J013350.3+303526.3 &   21.50 &   24.38 &    0.35 &    1.30 &      1\nl 
 D33J013341.6+303854.5 &   21.52 &   20.17 &    0.35 &    0.07 &      LP\nl 
 D33J013339.5+303649.0 &   21.55 & \nodata &    0.53 & \nodata &      LP\nl 
 D33J013408.7+303224.0 &   21.56 &   22.68 &    1.35 &    0.86 &       \nl 
 D33J013336.1+303835.9 &   21.59 & \nodata &    0.55 & \nodata &       \nl 
 D33J013357.4+303620.1 &   21.59 &   22.88 &    0.51 &    0.55 &       \nl 
 D33J013348.6+303842.2 &   21.62 & \nodata &    0.40 & \nodata &       \nl 
 D33J013410.5+303510.6 &   21.63 & \nodata &    0.67 & \nodata &       \nl 
 D33J013336.0+303801.1 &   21.64 &   22.50 &    0.48 &    0.28 &       \nl 
 D33J013346.2+303109.9 &   21.64 &   22.56 &    0.89 &    0.37 &       \nl 
 D33J013415.3+303808.7 &   21.64 &   22.28 &    0.57 &    0.19 &       \nl 
 D33J013345.3+303449.8 &   21.66 & \nodata &    0.81 & \nodata &       \nl 
 D33J013333.2+303516.9 &   21.68 &   23.18 &    1.71 &    1.72 &       \nl 
 D33J013402.4+303513.1 &   21.68 &   22.94 &    0.61 &    0.80 &       \nl 
 D33J013354.4+303210.2 &   21.69 &   21.72 &    0.33 &    0.26 &       \nl 
 D33J013412.2+303241.6 &   21.71 &   23.17 &    0.58 &    0.53 &      1\nl 
 D33J013335.0+303056.6 &   21.72 &   22.71 &    0.90 &    0.49 &       \nl 
 D33J013339.7+303003.9 &   21.72 &   23.53 &    0.66 &    0.83 &       \nl 
 D33J013342.3+303153.1 &   21.72 & \nodata &    0.72 & \nodata &       \nl 
 D33J013353.0+302904.1 &   21.76 &   23.12 &    0.75 &    0.97 &       \nl 
 D33J013404.5+303315.4 &   21.77 &   23.66 &    0.71 &    0.96 &      1\nl 
 D33J013409.4+303800.4 &   21.77 &   22.52 &    0.80 &    0.28 &       \nl 
 D33J013343.8+303116.7 &   21.80 & \nodata &    0.40 & \nodata &       \nl 
 D33J013346.7+303851.0 &   21.82 & \nodata &    0.51 & \nodata &       \nl 
 D33J013346.3+303556.1 &   21.86 &   22.92 &    0.61 &    0.49 &       \nl 
 D33J013333.5+303624.4 &   21.88 & \nodata &    0.77 & \nodata &       \nl 
 D33J013404.3+303018.5 &   21.89 &   23.33 &    0.73 &    0.58 &      LP\nl 
 D33J013408.2+303705.5 &   21.89 &   23.30 &    0.70 &    0.54 &       \nl 
 D33J013401.3+303501.1 &   21.90 &   23.23 &    0.44 &    1.12 &      LP\nl 
 D33J013409.8+303652.2 &   21.91 & \nodata &    1.39 & \nodata &       \nl 
 D33J013402.7+303656.7 &   21.92 &   22.90 &    0.91 &    0.46 &       \nl 
 D33J013408.0+303210.2 &   21.92 & \nodata &    0.48 & \nodata &       \nl 
 D33J013350.2+303712.6 &   21.96 & \nodata &    0.85 & \nodata &       \nl 
 D33J013337.1+303337.2 &   21.98 &   22.20 &    1.28 &    0.35 &       \nl 
 D33J013402.5+303256.5 &   21.98 & \nodata &    0.36 & \nodata &       \nl 
 D33J013334.3+303856.0 &   22.00 & \nodata &    1.76 & \nodata &       \nl 
 D33J013355.4+303259.4 &   22.00 &   23.27 &    0.97 &    0.75 &       \nl 
 D33J013338.2+303438.9 &   22.05 &   23.35 &    1.47 &    1.38 &       \nl 
 D33J013406.4+303042.0 &   22.05 &   23.47 &    0.48 &    0.50 &       \nl 
 D33J013406.9+303759.6 &   22.05 & \nodata &    0.60 & \nodata &       \nl 
 D33J013335.2+303653.1 &   22.06 &   23.55 &    1.02 &    0.84 &       \nl 
 D33J013403.2+303838.2 &   22.06 &   23.25 &    0.85 &    0.58 &       \nl 
 D33J013358.7+303450.5 &   22.07 & \nodata &    0.96 & \nodata &       \nl 
 D33J013342.7+303137.3 &   22.13 &   23.04 &    1.49 &    0.84 &       \nl 
 D33J013401.4+303753.8 &   22.14 & \nodata &    0.70 & \nodata &       \nl 
 D33J013405.9+303733.1 &   22.16 &   23.08 &    1.01 &    0.60 &       \nl 
 D33J013336.7+303836.0 &   22.21 &   23.17 &    0.51 &    0.51 &       \nl 
 D33J013353.8+303028.2 &   22.23 & \nodata &    1.45 & \nodata &       \nl 
 D33J013343.2+302903.3 &   22.25 & \nodata &    1.46 & \nodata &       \nl 
 D33J013338.4+303135.5 &   22.27 & \nodata &    0.71 & \nodata &       \nl 
 D33J013416.6+303300.2 &   22.27 & \nodata &    0.73 & \nodata &       \nl 
 D33J013339.3+303033.1 &   22.29 &   22.91 &    1.60 &    1.26 &      LP\nl 
 D33J013409.3+303357.7 &   22.30 &   23.22 &    1.39 &    0.90 &       \nl 
 D33J013344.3+303334.0 &   22.32 & \nodata &    1.10 & \nodata &       \nl 
 D33J013406.7+303128.6 &   22.33 &   23.92 &    1.07 &    0.94 &       \nl 
 D33J013355.9+303744.0 &   22.38 & \nodata &    1.42 & \nodata &       \nl 
 D33J013334.1+302956.5 &   22.41 & \nodata &    1.17 & \nodata &       \nl 
 D33J013414.8+303753.8 &   22.41 & \nodata &    1.05 & \nodata &       \nl 
 D33J013353.9+303635.6 &   22.44 &   24.46 &    1.07 &    1.10 &       \nl 
 D33J013346.8+303721.4 &   22.45 &   23.84 &    1.28 &    1.64 &       \nl 
 D33J013357.3+303811.5 &   22.45 & \nodata &    1.91 & \nodata &       \nl 
 D33J013412.4+303350.1 &   22.45 & \nodata &    1.34 & \nodata &       \nl 
 D33J013335.9+303644.6 &   22.49 & \nodata &    0.87 & \nodata &      LP\nl 
 D33J013349.4+303853.3 &   22.49 & \nodata &    1.56 & \nodata &       \nl 
 D33J013358.3+303055.1 &   22.50 &   23.72 &    0.74 &    0.69 &       \nl 
 D33J013359.0+303315.3 &   22.52 & \nodata &    1.68 & \nodata &       \nl 
 D33J013402.1+302946.3 &   22.53 &   23.50 &    0.96 &    0.41 &       \nl 
 D33J013415.0+303517.2 &   22.55 & \nodata &    1.89 & \nodata &       \nl 
 D33J013353.8+303145.5 &   22.56 & \nodata &    1.25 & \nodata &       \nl 
 D33J013338.3+303224.1 &   22.60 & \nodata &    0.88 & \nodata &       \nl 
 D33J013413.2+303251.1 &   22.60 &   23.29 &    0.74 &    0.65 &       \nl 
 D33J013404.0+303046.2 &   22.61 & \nodata &    0.94 & \nodata &       \nl 
 D33J013353.2+303237.0 &   22.62 & \nodata &    0.86 & \nodata &       \nl 
 D33J013404.5+303257.0 &   22.66 &   23.61 &    0.87 &    0.71 &      1\nl 
 D33J013335.4+303157.1 &   22.67 & \nodata &    0.98 & \nodata &       \nl 
 D33J013416.8+303214.9 &   22.68 & \nodata &    1.39 & \nodata &       \nl 
 D33J013415.9+302919.2 &   22.69 &   23.46 &    0.87 &    0.44 &      1\nl 
 D33J013347.3+303646.0 &   22.72 & \nodata &    1.19 & \nodata &       \nl 
 D33J013349.4+303423.0 &   22.77 &   23.76 &    1.26 &    0.81 &       \nl 
 D33J013350.6+303133.8 &   22.87 & \nodata &    1.30 & \nodata &       \nl 
 D33J013417.6+303649.9 &   22.89 & \nodata &    1.36 & \nodata &       \nl 
 D33J013411.4+303224.4 &   22.92 &   23.96 &    1.32 &    1.01 &       \nl 
 D33J013334.8+302930.4 &   22.93 & \nodata &    1.13 & \nodata &       \nl 
 D33J013338.8+303312.9 &   22.93 & \nodata &    1.61 & \nodata &       \nl 
 D33J013345.6+303704.8 &   22.96 &   23.60 &    1.28 &    0.68 &       \nl 
 D33J013355.3+303832.5 &   23.21 & \nodata &    2.07 & \nodata &       \nl 
 D33J013333.1+303756.6 &   23.23 & \nodata &    1.42 & \nodata &       \nl 
 D33J013413.3+303133.1 &   23.32 &   24.31 &    1.08 &    1.06 &       \nl 
 D33J013405.2+303653.9 &   23.43 & \nodata &    2.19 & \nodata &       \nl 
 D33J013331.4+303407.5 & \nodata & \nodata & \nodata & \nodata &       \nl 
 D33J013332.2+303016.9 & \nodata & \nodata & \nodata & \nodata &      LP\nl 
 D33J013333.8+303854.8 & \nodata &   23.05 & \nodata &    0.52 &       \nl 
 D33J013334.4+303011.0 & \nodata & \nodata & \nodata & \nodata &       \nl 
 D33J013334.8+303835.7 & \nodata & \nodata & \nodata & \nodata &      1\nl 
 D33J013335.0+302912.3 & \nodata & \nodata & \nodata & \nodata &       \nl 
 D33J013335.2+303406.0 & \nodata & \nodata & \nodata & \nodata &       \nl 
 D33J013335.4+303829.8 & \nodata & \nodata & \nodata & \nodata &       \nl 
 D33J013335.8+303038.4 & \nodata &   22.67 & \nodata &    0.25 &       \nl 
 D33J013336.0+303359.4 & \nodata & \nodata & \nodata & \nodata &       \nl 
 D33J013336.5+303111.8 & \nodata & \nodata & \nodata & \nodata &       \nl 
 D33J013336.5+303339.0 & \nodata & \nodata & \nodata & \nodata &       \nl 
 D33J013336.6+303154.3 & \nodata & \nodata & \nodata & \nodata &      LP\nl 
 D33J013336.7+303505.7 & \nodata & \nodata & \nodata & \nodata &       \nl 
 D33J013338.3+303739.2 & \nodata &   23.64 & \nodata &    0.77 &       \nl 
 D33J013339.1+303536.0 & \nodata & \nodata & \nodata & \nodata &       \nl 
 D33J013339.5+303824.5 & \nodata &   21.39 & \nodata &    0.30 &       \nl 
 D33J013339.7+303827.5 & \nodata & \nodata & \nodata & \nodata &       \nl 
 D33J013339.8+303451.9 & \nodata & \nodata & \nodata & \nodata &      1\nl 
 D33J013341.1+303717.0 & \nodata & \nodata & \nodata & \nodata &       \nl 
 D33J013342.2+303747.0 & \nodata &   20.43 & \nodata &    0.93 &      LP\nl 
 D33J013342.5+303718.4 & \nodata & \nodata & \nodata & \nodata &       \nl 
 D33J013342.6+303300.5 & \nodata & \nodata & \nodata & \nodata &       \nl 
 D33J013342.7+303146.5 & \nodata & \nodata & \nodata & \nodata &      LP\nl 
 D33J013342.7+303805.0 & \nodata & \nodata & \nodata & \nodata &       \nl 
 D33J013342.9+303234.4 & \nodata & \nodata & \nodata & \nodata &      1\nl 
 D33J013343.9+303735.8 & \nodata &   22.41 & \nodata &    0.28 &       \nl 
 D33J013345.2+303437.0 & \nodata & \nodata & \nodata & \nodata &       \nl 
 D33J013345.2+303547.4 & \nodata & \nodata & \nodata & \nodata &       \nl 
 D33J013345.4+303823.1 & \nodata & \nodata & \nodata & \nodata &       \nl 
 D33J013345.9+303620.0 & \nodata & \nodata & \nodata & \nodata &       \nl 
 D33J013346.4+303822.7 & \nodata &   21.60 & \nodata &    0.20 &       \nl 
 D33J013346.7+303902.5 & \nodata & \nodata & \nodata & \nodata &       \nl 
 D33J013346.9+303522.6 & \nodata & \nodata & \nodata & \nodata &       \nl 
 D33J013348.9+303530.2 & \nodata & \nodata & \nodata & \nodata &       \nl 
 D33J013349.5+303822.4 & \nodata &   22.66 & \nodata &    0.28 &       \nl 
 D33J013349.9+303838.6 & \nodata & \nodata & \nodata & \nodata &      LP\nl 
 D33J013349.9+303852.2 & \nodata &   19.14 & \nodata &    0.07 &      LP\nl 
 D33J013350.5+303654.8 & \nodata &   24.55 & \nodata &    1.69 &       \nl 
 D33J013350.6+303842.5 & \nodata &   19.45 & \nodata &    0.05 &       \nl 
 D33J013350.7+303158.4 & \nodata & \nodata & \nodata & \nodata &       \nl 
 D33J013351.8+303849.1 & \nodata &   18.05 & \nodata &    0.07 &       \nl 
 D33J013351.9+303529.1 & \nodata & \nodata & \nodata & \nodata &       \nl 
 D33J013352.0+303152.8 & \nodata &   22.32 & \nodata &    0.81 &      LP\nl 
 D33J013352.5+303903.1 & \nodata & \nodata & \nodata & \nodata &       \nl 
 D33J013352.8+303602.2 & \nodata & \nodata & \nodata & \nodata &       \nl 
 D33J013353.0+303200.9 & \nodata & \nodata & \nodata & \nodata &       \nl 
 D33J013353.0+303759.3 & \nodata & \nodata & \nodata & \nodata &      LP\nl 
 D33J013353.0+303815.2 & \nodata & \nodata & \nodata & \nodata &       \nl 
 D33J013353.5+303823.9 & \nodata &   18.52 & \nodata &    0.10 &      LP\nl 
 D33J013354.2+303721.2 & \nodata & \nodata & \nodata & \nodata &      LP\nl 
 D33J013354.8+303815.1 & \nodata & \nodata & \nodata & \nodata &       \nl 
 D33J013354.9+303450.6 & \nodata & \nodata & \nodata & \nodata &       \nl 
 D33J013355.5+303400.0 & \nodata & \nodata & \nodata & \nodata &       \nl 
 D33J013356.0+303834.7 & \nodata &   20.05 & \nodata &    0.04 &      1\nl 
 D33J013356.5+303604.0 & \nodata &   22.84 & \nodata &    0.28 &       \nl 
 D33J013356.5+303812.1 & \nodata & \nodata & \nodata & \nodata &       \nl 
 D33J013356.8+303430.2 & \nodata & \nodata & \nodata & \nodata &      LP\nl 
 D33J013356.9+302949.4 & \nodata & \nodata & \nodata & \nodata &       \nl 
 D33J013357.0+303818.0 & \nodata &   21.14 & \nodata &    0.22 &      1\nl 
 D33J013357.1+303559.3 & \nodata & \nodata & \nodata & \nodata &       \nl 
 D33J013357.2+303837.9 & \nodata &   21.42 & \nodata &    0.15 &       \nl 
 D33J013357.6+303844.0 & \nodata & \nodata & \nodata & \nodata & 2    LP\nl 
 D33J013357.8+303717.9 & \nodata &   20.84 & \nodata &    0.16 &      1\nl 
 D33J013358.1+303235.2 & \nodata & \nodata & \nodata & \nodata &       \nl 
 D33J013359.7+303753.2 & \nodata & \nodata & \nodata & \nodata &       \nl 
 D33J013401.7+303601.1 & \nodata &   18.34 & \nodata &    0.02 &      LP\nl 
 D33J013401.9+303908.3 & \nodata & \nodata & \nodata & \nodata &       \nl 
 D33J013402.0+303901.2 & \nodata &   22.66 & \nodata &    0.54 &       \nl 
 D33J013402.1+303836.0 & \nodata & \nodata & \nodata & \nodata &       \nl 
 D33J013402.9+303316.8 & \nodata &   22.17 & \nodata &    0.26 &       \nl 
 D33J013402.9+303754.2 & \nodata &   18.84 & \nodata &    0.04 &      LP\nl 
 D33J013403.4+303649.3 & \nodata &   18.39 & \nodata &    0.06 &       \nl 
 D33J013405.4+303719.2 & \nodata &   20.62 & \nodata &    0.21 &      1\nl 
 D33J013405.6+303119.6 & \nodata &   23.43 & \nodata &    0.40 &       \nl 
 D33J013407.8+303334.5 & \nodata &   22.66 & \nodata &    0.47 &       \nl 
 D33J013408.5+303631.5 & \nodata & \nodata & \nodata & \nodata &       \nl 
 D33J013409.3+303414.4 & \nodata &   18.77 & \nodata &    0.07 &      LP\nl 
 D33J013409.4+303706.2 & \nodata & \nodata & \nodata & \nodata &      1\nl 
 D33J013409.6+303908.0 & \nodata &   21.34 & \nodata &    0.15 &      1\nl 
 D33J013409.7+303829.9 & \nodata &   23.70 & \nodata &    0.92 &       \nl 
 D33J013410.3+303710.0 & \nodata & \nodata & \nodata & \nodata &       \nl 
 D33J013410.9+303437.5 & \nodata &   16.09 & \nodata &    0.08 &      1\nl 
 D33J013411.1+303659.5 & \nodata & \nodata & \nodata & \nodata &      LP\nl 
 D33J013413.8+303337.3 & \nodata & \nodata & \nodata & \nodata &       \nl 
 D33J013415.2+303207.5 & \nodata & \nodata & \nodata & \nodata &       \nl 
 D33J013415.3+303404.6 & \nodata & \nodata & \nodata & \nodata &      LP\nl 
 D33J013415.3+303804.3 & \nodata & \nodata & \nodata & \nodata &       \nl 
 D33J013416.1+303808.0 & \nodata &   18.97 & \nodata &    0.04 &       \nl 
 D33J013416.3+303712.3 & \nodata & \nodata & \nodata & \nodata &      LP\nl 
\enddata
\tablecomments{(1) Variables identified by Macri et al. (2001a)\\
(2) Variables identified by Mochejska et al. (2001a)}
\label{tab:misc}
\end{planotable}
\end{small}

\begin{small}
\tablenum{10}
\begin{planotable}{ccccc}
\tablewidth{0pc}
\tablecaption{\sc Light Curves of Miscellaneous Variables in M33B}
\tablehead{ \colhead{Name} & \colhead{Filter} &
\colhead{HJD$-$2451000} & \colhead{mag} & \colhead{$\sigma_{mag}$} }
\startdata
D33J013352.4+303840.2 & B & 452.7982   &   19.995  &  0.009\\
                      & B & 452.9166   &   19.987  &  0.012\\
                      & B & 454.6968   &   20.040  &  0.011\\ 
                      & B & 454.7299   &   20.040  &  0.009\\
\enddata
\tablecomments{Table 10 is available in its entirety in the electronic
version of the Astronomical Journal. A portion is shown here for
guidance regarding its form and content.}
\label{tab:lc_misc}
\end{planotable}
\end{small}

\end{document}